\documentclass{aa}  

\usepackage{graphicx}
\usepackage{txfonts}
\usepackage[]{natbib} 
\begin{document} 

\title{Gaseous nebulae and massive stars in the giant \ion{H}{I} ring in Leo}

\author{Edvige Corbelli \inst{1}
\and Filippo Mannucci\inst{1}
\and David Thilker\inst{2}
\and Giovanni Cresci\inst{1} 
\and Giacomo Venturi \inst{3,1}
 }

\institute {INAF-Osservatorio di Arcetri, Largo E. Fermi 5, 50125 Firenze, Italy \\
 \email{edvige.corbelli@inaf.it} 
 \and Department of Physics and Astronomy, The Johns Hopkins University, Baltimore, MD, USA
  \and Instituto de Astrof\'{i}sica, Facultad de F\'{i}sica, Pontificia Universidad Cat\'{o}lica de Chile, Casilla 306, Santiago 22, Chile\\ 
  }

\date{Received ....; accepted ......}

 \abstract
  {Chemical abundances in the Leo ring, the largest  \ion{H}{I} cloud in the local Universe, have recently been determined to be close or above solar \citep{2021ApJ...908L..39C}, incompatible with a previously claimed primordial origin of the ring. The gas, pre-enriched in a galactic disk and tidally stripped, did not manage to form stars very efficiently in intergalactic space.
  }  
{Using H$\alpha$ emission and a multi wavelengths analysis of  its extremely faint optical counterpart we investigate the process of star formation and the slow building up of  a stellar population.
 }  
{We map nebular lines in 3 dense \ion{H}{I} clumps of the Leo ring and complement these data with archival stellar continuum observations and population synthesis models. 
}
 {A sparse population of stars is detected  in the main body of the ring, with individual young stars as massive as O7-types powering some \ion{H}{II} region. The average star formation rate density in the ring is of order of 10$^{-5}$~$M_\odot$~yr$^{-1}$~kpc$^{-2}$ and proceeds with local bursts a few hundred parsecs in size, where loose stellar associations of 500 -- 1000~$M_\odot$ occasionally host massive outliers.  The  far ultraviolet -to-H$\alpha$ emission ratio in nebular regions implies recent stellar bursts, from 2 to 7~Myr ago. The relation between the local \ion{H}{I} gas density and the star formation rate in the ring is similar to what is found in dwarfs and outer disks with gas depletion times  as long as 100~Gyrs.  We find a candidate planetary nebula in a compact  and faint H$\alpha$ region with [OIII]/H$\alpha$ line enhancement which is consistent with the estimated mean stellar surface brightness of the ring. The presence of 1~kpc partial ring emitting weak  H$\alpha$ lines  around the brightest  and youngest \ion{H}{II} region, in a gas clump towards M\,96, suggests that local shocks might be the triggers of future star forming sites.
   }
 {}


\maketitle

\section{Introduction}

The origin of the largest extragalactic neutral gas cloud in the local Universe ($D\le 20$~Mpc), the Leo ring  \citep{1983ApJ...273L...1S,1985ApJ...288L..33S}, has been a long standing mystery. Lacking a pervasive optical counterpart, the ring has been proposed as a rare candidate primordial cloud dating to the time of the Leo I group formation \citep{1989AJ.....97..666S,2003ApJ...591..185S}. The Galaxy Evolution Explorer (GALEX) detection of ultraviolet (UV) continuum light in the direction of a few \ion{H}{I} clumps  of the ring, has opened the possibility of localized star formation between 0.1 and 1 Gyrs ago \citep{2009Natur.457..990T}.  A low level of metal enrichment inferred by the UV color ratios and by a tentative measurement of the metallicity  from QSO absorptions in nearby sightlines  \citep{2014ApJ...790...64R} favored the ring primordial origin hypothesis. Massive \ion{H}{I} clouds without extended star formation do not fit the current models of galaxy formation based on  $\Lambda$CDM cosmology and the paucity of such clouds suggests that local analogues have either dispersed, become ionized  or formed a stellar population on Gyrs timescales. The ring might have condensed from warm intergalactic medium and may be fueling M\,96.  Neutral gas cloudlets have in fact been found from the bulk of the ring towards this galaxy, with no extended diffuse stellar streams indicative of tidal interactions. If feeding of star formation in galaxies takes place from cooling  and accretion of intergalactic gas it is conceivable  that neutral condensation form for a short time as the intergalactic  gas enters the potential well of galaxies  and galaxy groups \citep{2009Natur.457..451D,2009MNRAS.395..160K,2014ApJ...783...45S}.  

The Leo ring on the other hand might have formed out of enriched gas, tidally stripped during a galaxy-galaxy head-on collision \citep{1951ApJ...113..413S,1985ApJ...288..535R}  or other tidal event such as an encounter of a low surface brightness galaxy with the group \citep{2005MNRAS.357L..21B}.  A close analogue to the Leo ring, the giant collisional ring around NGC\,5291, of similar size (200~kpc) but at a larger distance ($D\simeq 50$~Mpc) and more massive  \citep{1979MNRAS.188..285L,1997AJ....114.1427M} is experiencing a vigorous star formation with giant HII regions which are sites of dwarf galaxy formation \citep{1998A&A...333..813D,2007Sci...316.1166B,2007A&A...467...93B}. 
Opposite to NGC\,5291 and to other collisional ring galaxies such as  UGC\,7069 or the Cartwheel galaxy \citep{2008MNRAS.386L..38G,1996AJ....112.1868S}, the Leo ring is much more quiescent. A close analogue, a quiescent ring about half the size of the Leo ring, has recently been discovered around a massive quenched galaxy AGC\,203001 \citep{2020MNRAS.492....1B}. For many decades there have been no identified HII regions in the Leo ring which could provide metal lines and  chemical abundances, key ingredients for solving the mystery of its origin.  Taking advantage of MUSE (Multi Unit Spectroscopic Explorer) at ESO Very Large Telescope (VLT), our team has acquired high sensitivity integral field maps across the area of  3 \ion{H}{I} clumps in the Leo ring and found ionized nebulae with hydrogen and heavy element emission lines \citep[][hereafter Paper I]{2021ApJ...908L..39C}.  Chemical abundances in the ring are close or above solar values, a finding which does not support the ring primordial origin hypothesis. On the contrary, a high metallicity gas and a very weak stellar counterpart  can only  be interpreted as the footprint of pre-enriched gas  removed  from a galaxy disk. The M\,96 group is a compact group and the nearest one where both elliptical and spiral galaxies are found, as well as lenticular and dwarf irregular galaxies. A viable model for the Leo ring formation involving group members such as NGC\,3384 and M\,96  \citep{2010ApJ...717L.143M}  implies a formation time about 1~Gyr ago. Numerical simulations are  consistent with the metal rich gas being removed from  galaxy disks and placed in ring-like shape. The alignment between the spins of circumnuclear stellar and gaseous subsystems of NGC\,3384 and that  of the intergalactic \ion{H}{I} ring \citep{2003ApJ...591..185S} can  support the collisional hypothesis. 
The M\,96 young outer ring, very bright and dominant structure at 21-cm  \citep{1989ApJ...343...94S}  could be gas accreting after it had been ejected from NGC\,3384  \citep{2014ApJ...791...38W}. 

The main open question on the Leo ring is now why such extended structure of tidal origin did not experience strong compressions and extended stellar bursts. But before addressing this question it is important to characterize the process of star formation in the ring. In this paper we use GALEX UV continuum archival data and Hubble Space Telescope (HST) archival  imaging together with MUSE spectroscopic data  to trace  the star formation process and the massive stellar population in such peculiar environment. A low level on-going star formation can  slowly build up diffuse faint dwarf galaxies and we might capture a possible  process  of galaxy formation in action \citep{2018MNRAS.476.4565B}.
Deep optical imaging \citep{2009AJ....138..338S, 2015PKAS...30..517K,2018ApJ...863L...7M,2018ApJ...868...96C} reported diffuse faint dwarf galaxies close to the ring. There have been 3 spectroscopically confirmed  low luminosity diffuse galaxies  identified  at the ring edge \citep{2009AJ....138..338S} which might have resulted from sporadic  star formation episodes. Several  ultra diffuse galaxy candidates lie further out and are  close in projection  to the ring \citep{2018A&A...615A.105M}.  Determining the origin and fate of intergalactic clouds is therefore important as we develop a deeper understanding of the accretion and feedback processes that shape galaxy evolution and of the formation of diffuse dwarf galaxies. 

Our knowledge of the processes in the interstellar medium (ISM) that favor the birth of stars, is based on Galactic studies, where self-gravitating molecular clouds form, or are relative to shocked gas recently removed from  galaxies through tidal encounters \citep[e.g.][]{2004A&A...426..471L,2000Natur.403..867B}. Examining how stars  form in more quiescent intergalactic clouds, is  very relevant as they may be channels for polluting the intergalactic medium with metals, and can shed light on new modes of star formation, unveiling timescale and triggers of gas overdensities when physical conditions are different than in galaxies. With a neutral gas mass equivalent to that of a star forming galaxy, M$_{\rm{\ion{H}{I}} }\simeq 2\times 10^9$~$M_\odot$, the Leo ring provides a unique opportunity to investigate star formation in a metal rich but low density environment.  Local star formation rate (SFR) surface density ($\Sigma_{SFR}$) values of several 10$^{-4}$~$M_\odot$ yr$^{-1}$~kpc$^{-2}$ have been  found by \citet{2009Natur.457..990T} placing the Leo ring in the low extreme range of star forming objects.  The lack of a pervasive stellar component makes the local gravitational field weaker and the gas physical conditions more similar to those in outer disks of spiral galaxies and gas rich dwarfs. However,  other differences, such as the angular momentum and the large distance to a bright star forming disks make the Leo ring a unique gaseous environment for studying star formation.

The plan of the paper is the following. In Section~2 we summarize the main results of previous works on the Leo ring related to the subject of this paper, such as the detection of \ion{H}{I} clumps, of the associated UV emission and of metal rich  ionized nebulae, and describe the datasets used.  In Section~3 we  derive characteristic physical quantities of the newly discovered \ion{H}{II} regions and present the first  Planetary Nebula candidate in the ring. A comparison of star and stellar synthesis models with UV, optical, and H$\alpha$ data in Section~4 defines the properties of massive stars and of the stellar population in and around the \ion{H}{II} regions. The current star formation rates  across the ring are analyzed in Section~5.  Section~6 summarizes and concludes.  

Throughout this paper we assume a distance to  the Leo ring of 10~Mpc, as for M\,96 and M\,105. This implies that an angular separation of 1\arcsec\   corresponds to a spatial scale of 48.5~pc.

\section{Past surveys and the data}

The Leo ring has been subject of a large number of observations and studies. Here we summarize the most relevant results for the subject of this paper. We then describe the data and  extinction corrections used in this work. 

\subsection{The cold gas}

The low  \ion{H}{I} column density contours of the ring, as mapped by Arecibo with the 3.3\arcmin\ half-power beam   \citep{1989AJ.....97..666S}, are plotted in magenta in the left panel of Figure~1. Radial velocities, between 860 and 1060~km~s$^{-1}$, indicates ordered rotational motion of the whole ring around the three galaxies at the center, shown in the background optical image. The bulk of the \ion{H}{I}  gas in the ring is on the south and west side, especially between M\,96 (to the south) and NGC\,3384/M\,105, while to the north and east side a few discrete and more diffuse clouds are found. The rotational velocities imply a dynamical time of order of 4~Gyrs and a dynamical mass inside the ring of 6$\times 10^{11}$~$M_\odot$ \citep{1985ApJ...288L..33S}, only a factor 2 higher than that inferred for the central galaxies. 
Higher resolution VLA (Very Large Array) maps have been carried out in D-configuration \citep{1986AJ.....91...13S} with an effective beam of 45\arcsec\ centered on the southern part of the cloud. The \ion{H}{I} contours of the  interferometer image  (in yellow in Figure~1)  revealed the presence of gas clumps  with masses between 1 and 3.5$\times 10^7$~$M_\odot$, and surface areas between 3 and 10~kpc$^2$.  At this spatial resolution the peak column densities reach 4$\times 10^{20}$~cm$^{-2}$ and the most massive clumps appear as distinct virialized entities \citep{1986AJ.....91...13S}. Some show internal rotation and a spatial elongation which suggests a disk like geometry.  The velocity field becomes complex in the extension pointing south, towards M\,96, made of distinct cloudlets  with masses of order of 10$^7$~$M_\odot$.  

At the location of a few \ion{H}{I} peaks CO searches  have been carried out using the Five College Radio Astronomy Observatory telescope \citep{1989AJ.....97..666S}  providing an upper limit of 0.8~K~km~s$^{-1}$. These measures exclude  large molecular-to-atomic gas ratios in \ion{H}{I} clumps since for a galactic CO-to-H$_2$ conversion factor they imply a column density ratio N(H$_2$)/N(\ion{H}{I})$<$0.5.  

\begin{figure*} 
\includegraphics [width=7.5 cm, angle=0 ]{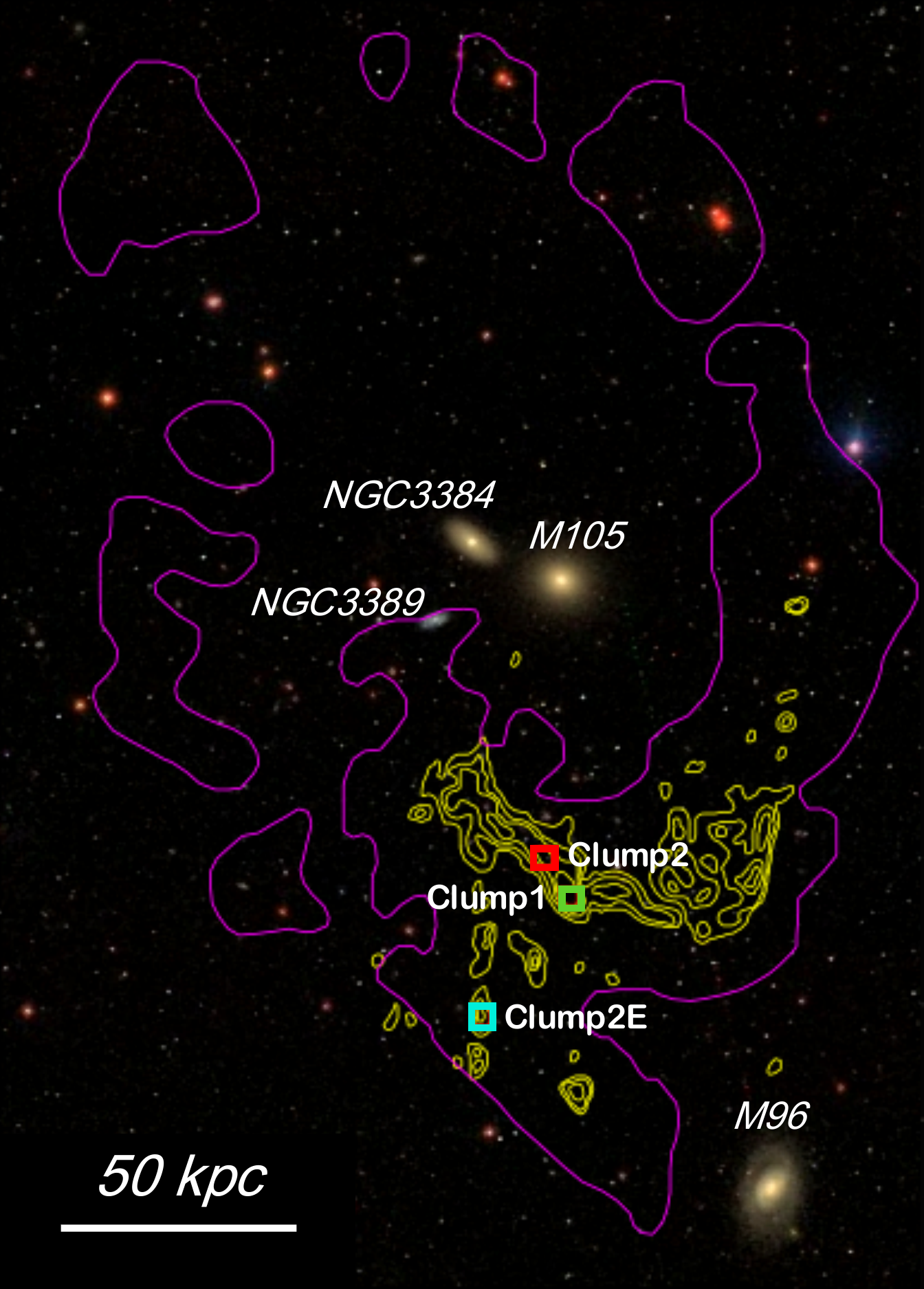}
\includegraphics [width=10.7 cm]{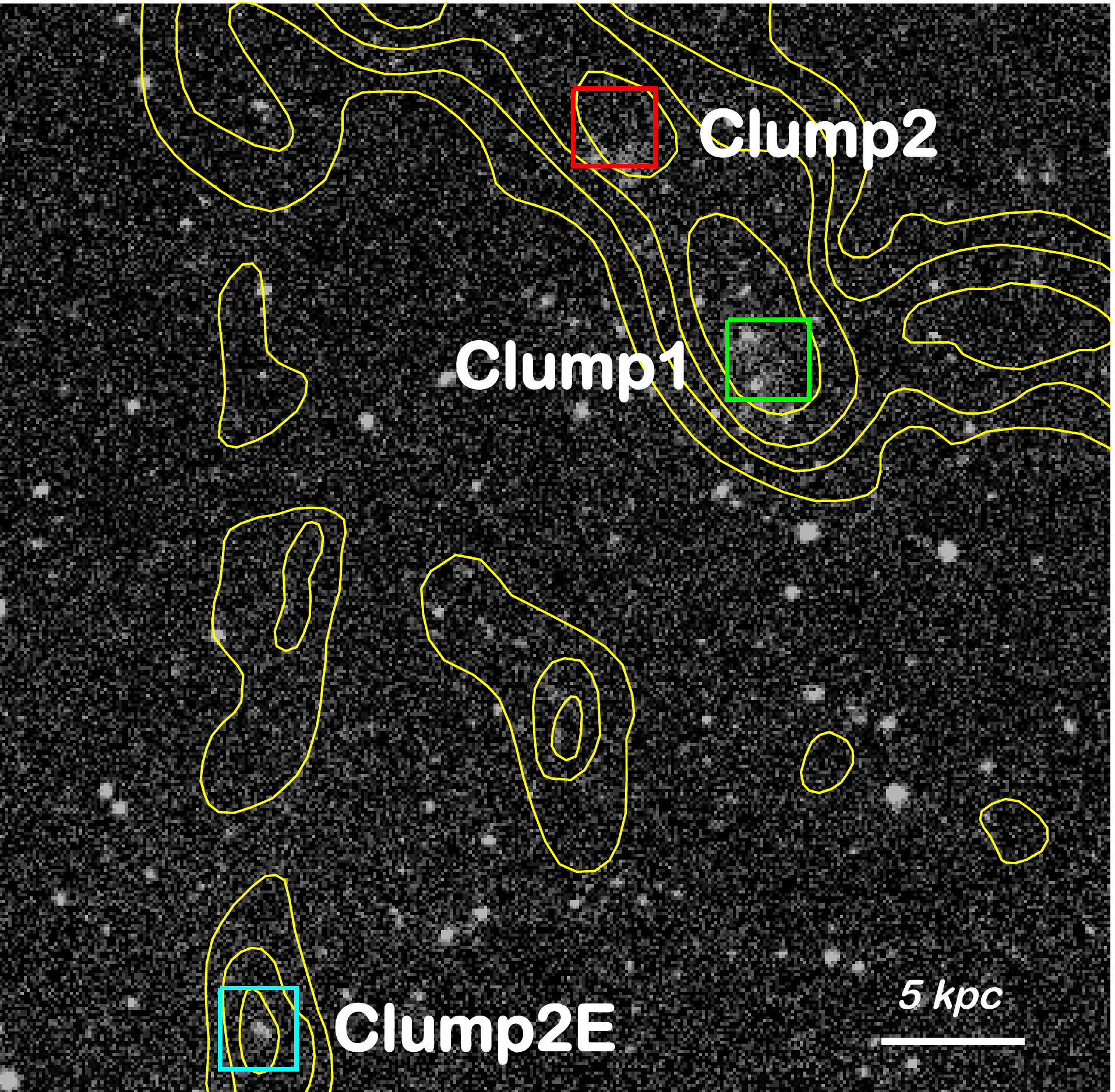}
\caption{The optical image of the M96 group in the background (SDSS color image) is shown in the left panel with \ion{H}{I} contours of the Leo ring.  In magenta the Arecibo contour at N$_{HI}=2\times 10^{18}$~cm$^{-2}$, in yellow the VLA \ion{H}{I} contours of  the southern part of the ring as described by \citet{1986AJ.....91...13S}. Square symbols indicate the positions of the 3 \ion{H}{I} clumps observed with MUSE: Clump1, Clump2 and Clump2E.  In the right panel an enlargement of the 3 \ion{H}{I} clumps shows the coverage of the 8.5~kpc$^2$ MUSE fields overlaied on the far UV-GALEX  image.  
}
\label{fig:fields}
\end{figure*}

 \subsection{Stellar footprints in and around the ring}

For many years the Leo ring was detected only via \ion{H}{I} emission. Deep optical surveys suggested the complete absence of  an extended stellar population in the ring or interior to it  with $\mu_B>30$~mag~arcsec$^{-2}$ \citep{2014ApJ...791...38W,1985AJ.....90..450P,1985MNRAS.213..111K}. A narrow band OIII line survey of the high \ion{H}{I} column density 1122~arcmin$^{2}$ region of the ring  \citep{2003A&A...405..803C} has been carried out to identify planetary nebulae (PNe) down to a limiting [OIII] magnitude $m_{5007}=27.49$.  The  absence  of  spectroscopically confirmed  PN  candidate  gives  more  stringent  upper  limits to  the  surface  brightness  of the  old  stellar  population  associated  with  the  Leo cloud, with  $\mu_B>32.8$~mag~arcsec$^{-2}$.
The lack of an optical counterpart and of a significant intragroup diffuse starlight  is expected for a loose group with rare encounters,  but it does not explain the persistence of a gaseous ring forming very few stars. Given the short group crossing time and the fact that gas in galaxy-galaxy encounters tend to shock, lose angular momentum sinking to subkpc scale, it is unexpected that the bulk of the ring gas  is apparently stable and long lasting.
 
The first wide-field ultraviolet imaging of the Leo ring by GALEX, partially sampling the overall ring, revealed localized kpc size over densities of sources correlated with \ion{H}{I} gas peaks. Although there might be contamination by background objects,  \citet{2009Natur.457..990T}  attributed  the UV emission of sources  in close  projection to  \ion{H}{I} gas clumps, Clump1, Clump2, and Clump2E,  to localized events of massive star formation. These clumps are shown in the right panel of Figure~1.  The UV and optical colors favored  ages  of order of 0.1--1~Gyrs with  metallicities of order 0.02--0.2 Z$_\odot$.  Two of these regions, Clump1 and Clump2 lie in the southern part of the main body of the ring  which hosts most of the ring gas mass and has the highest gas mass surface density. The third region, Clump2E, is in the filament connecting the ring to M\,96. The UV emission presents several peaks in Clump1 and Clump2, typical of an extended star forming region, while it has a more compact morphology in Clump2E.  Deep optical imaging of the area has revealed some patchy light with blue colors, with $B-V$= 0.0$\pm$0.1,  0.2$\pm$0.2, and 0.1$\pm$0.2 for Clump1, Clump2, and Clump2E respectively \citep{2014ApJ...791...38W}.
Wide field-of-view optical images of the ring region  \citep{2018ApJ...863L...7M} have more recently revealed a localized very faint diffuse blue  optical counterpart (with $\mu_B$ =28.8~mag arcsec$^{-2}$) and some new UV knots  in one of the condensations connecting the \ion{H}{I} ring to the M\,96 galaxy. Colors indicate a post-burst stellar population 200-600~Myr old at different location in the ring  than  that  inferred by  \citet{2009Natur.457..990T}. Diffuse dwarf galaxies might  form continuously from ring material but  they  might also result from a pre-existing gas rich low surface brightness galaxy undergoing  tidal interaction with the group.

\subsection{Optical spectroscopy and the detection of nebular lines}

The first  spectroscopic detection of ionized hydrogen and nebular metal lines in the proximity of \ion{H}{I} gas clumps of the Leo ring have been presented in Paper I. The integral field spectrograph  MUSE  at VLT  has been centered at 3 \ion{H}{I} peak locations close to the  UV emission  detected by \citet{2009Natur.457..990T}. The 3 MUSE fields are shown in the right panel of Figure~\ref{fig:fields}  overalayed to the VLA \ion{H}{I} contours and to the GALEX far UV (FUV) image. Details of these observations are described in Paper I. The final dataset comprises 3 data cubes, one per clump,  covering a FoV slightly larger than 1~arcmin$^2$. Each spectrum spans the wavelength range 4600 - 9350~\AA\ with a spatial resolution given by the seeing, of  the order of 1\arcsec . 

The cubes covers completely 2 of the 3 ultraviolet-bright  regions  listed by \citet{2009Natur.457..990T}. The  UV region close to Clump2, is only partially covered by our observations. Five gaseous nebulae have been detected in H$\alpha$ and in at least one metal line: three are in the Clump1 field and two in the Clump2E field. 
We do not detect  nebular lines in Clump~2 and have upper limits of the order of 0.9 and 0.5$\times 10^{-17}$~erg~s$^{-1}$~cm$^{-2}$ in the blue and red parts of the spectrum respectively when averaging spectra in circular apertures with radius 1.2\arcsec . This clump is the reddest of the three clumps observed, having the largest values of  UV and optical  colors \citep{2009Natur.457..990T,2014ApJ...791...38W}. We cannot exclude some very weak H$\alpha$ line from regions  not coincident with the FUV peaks but close to them, towards the southern boundary of the MUSE field. Unfortunately the  southernmost side of the FUV emission  overlaps with the edge of the MUSE field and precludes any definitive conclusion. 

In Paper I we have shown the H$\alpha$ images of the five detected nebular regions  and analyzed the metal content of the four of them which have metal line ratios compatible with those observed in \ion{H}{II} regions. The fifth nebular regions, C1c, the faintest detected, has anomalous line ratios and is described in detail in the next Section. For the two \ion{H}{II} regions in Clump1, C1a and C1b,  we find chemical abundances  0.2 dex below solar values: for  the two \ion{H}{II} regions in Clump2E, C2Ea and C2Eb, the metallicity is well above solar. 

\subsection{GALEX and HST  archival data}

Since the analysis by \citet{2009Natur.457..990T} a newer GALEX data release with much deeper and dedicated observations of the entire Leo ring has been made available. The dataset we use is a deep coadd of all GALEX imaging in the far and near UV (FUV and NUV) that overlaps the ring.  This coadd reaches a minimum depth of ~15 ks [~9 ks] in NUV [FUV] over the regions in which \ion{H}{I} is detected, but has irregularly shaped patches reaching nearly twice this depth.  Clumps 1, 2 are in a region of ~24 ks [16 ks] in NUV [FUV], whereas Clump 2E coverage is slightly shallower at ~22 ks [11 ks].  The exposure time for Clumps 1, 2 yields a 5-sigma point source limit of  ~27.5 ABmag [25.8] in NUV [FUV].  At this depth, confusion from the background of distant galaxies is significant limitation on the NUV dataset, and thus we use the FUV image primarily (except for UV color measurements).   Galactic cirrus is also prominent across the GALEX coadd, complicating the association of diffuse UV features with the Leo Ring.

Owing to the low stellar density,  Leo ring  HST images   and color magnitude diagrams can help us resolving individual bright stars, identify  compact star clusters, and the age of the population. There is a serendipitous archival HST-Advanced Camera for Surveys (ACS) coverage  of the most prominent \ion{H}{I} clump, Clump1, obtained as pure parallel imaging in V, i, z bands (F606W, F775W, F850LP) during June 2011 as part of program PID 12286, led by H. Yan.  Unfortunately, no blue HST filter coverage was obtained at the time of these serendipitous observations.   Clump 2 and Clump 2E have never been  imaged by HST. 

We analyzed the archival ACS/WFC HST data using DOLPHOT to detect and measure point- and point-like sources in the field covering Clump1. The parameters for DOLPHOT were set to standard choices established for nearby galaxies by several large surveys (e.g. PHAT, LEGUS for which photometry was described by \citet{2014ApJS..215....9W,2018ApJS..235...23S}).  In our particular case, we photometered stars on a set of 11 individual exposures (three F606W, four F775W, four F850LP).  The summed integration time for this dataset was 2892s, 3908s, 3887s, respectively in order of the bands listed above.  PSF-fitting magnitudes were computed in the Vega system using zeropoints supplied in the image headers by the STScI pipeline. The 5-sigma point source limiting magnitude is 28.4, 27.5, 26.7 in F606W, F775W, and F850LP, respectively. We  detect only very sparse associations (no bright clusters or dwarf-like  galaxies) and a highly scattered distribution of stars which are described in Section~4.

\subsection{Extinction corrections}

The optical extinction inferred from fits to metal line ratios is low (Paper I), with $A_V\lessapprox 1$~mag. In Table~1 of Paper I we list $A_V$ for \ion{H}{II} regions using apertures with radii $R_{ap}$=1.2\arcsec. We consider these appropriate for extinction corrections in the core of the nebulae i.e. for $R_{ap}\le$1.2\arcsec . In this paper we use a variety of apertures  and we have computed extinction as we vary the aperture size using the Balmer decrement and the $R_V$=3.1 Milky Way extinction curve of \citet{2001ApJ...548..296W}.  For $R_{ap}$ between 1.5\arcsec\ and 3.5\arcsec\  we find $A_V$ values of 0.4$\pm$0.1 for C1a and of 0.6$\pm$0.2 for C2Ea. Balmer decrements for C1b and C2Eb cannot  be recovered in all apertures due to their intrinsic faint emission lines, with H$\beta$ line often below the detection threshold or with large uncertainties. The increase of the Balmer decrement extinction as the aperture size increases for C2Ea is due to faint H$\alpha$ emitters in the proximity of this HII region (see later in this Section). Therefore we use $A_V$=0.4~mag for emission line extinction throughout this paper. This extinction includes the minimal foreground extinction $A_V$=0.074~mag in the direction of the Leo ring \citep{1998ApJ...500..525S}. 

For the stellar UV or optical emission we use $A_V$=0.2~mag ($A_{FUV}$=0.5~mag), a factor 2 lower extinction than for emission lines in the nebulae due to additional absorption of Balmer lines by local dust and in agreement with the lower obscuration of stellar continuum observed and modeled around HII regions \citep{2001PASP..113.1449C}. Only for the optical emission from the center of the nebulae we consider extinction in the range given by $A_V$=0.4 and the values listed in Table~1 of Paper I.

In this paper we use the extinction corrected H$\alpha$ emission to estimate the ionizing photon rate emitted by stars or stellar clusters in the ionized nebulae. Dust in the nebulae can however absorb directly part of the ionizing photons which contribute to dust heating. The fraction of ionizing photons lost by this process depend on the dust destruction and radiation field inside the nebulae, on the dust and complex polycyclic aromatic hydrocarbon abundance in the ring which are hard to quantify  \citep{2001AJ....122.1788I,2003ApJ...583..727D,2009ApJ...703.1672K}. Given the estimated low radiation field in the C1a nebulae (see Section 3) however, we expect this fraction to be lower than in galaxies of similar metallicities. In addition, ionizing photons might leak out of the \ion{H}{II} regions. Leakage increase as  dust grains are destroyed in the \ion{H}{II} regions and as the regions evolve \citep{2019ApJ...883..102K}. The fraction of ionizing photons leaking out of the nebulae can be of the order of 50$\%$ after 5~Myr.
We don't explicitly take into account the fraction of ionizing photons directly absorbed by dust or leaking out of the nebulae  although we mention that the estimates ionizing photon rate might likely be  higher than what we quote due to these photon losses.

\section{\ion{H}{II} regions versus Planetary Nebulae}

Using apertures with radii of 1.2\arcsec\ and 2.4\arcsec\  we measure line flux ratios in the ionized nebulae from MUSE data. The smaller aperture is used to maximize the signal to noise of weak lines, being its radius of the order of the seeing, and the larger one covers most of the extent of the H$\alpha$ emission, with some strong line still above the threshold.  In Table~1 we show the position and integrated intensities of Gaussian fits to lines whose peaks are higher than 3 times the spectrum rms in the large aperture. Data relative to the small aperture  are shown in Paper I, except for C1c whose line fluxes in the 1.2\arcsec\  aperture are 50$\%$ and 65$\%$ lower for [OIII] and H$\alpha$ respectively than in the wider aperture. 
The upper limits in Table~1 are inferred using 3$\times$ rms of the spectra at the expected wavelengths and the full spectral extent of the line, typically 8~$\AA$.

\begin{figure} 
 \includegraphics[width=9 cm] {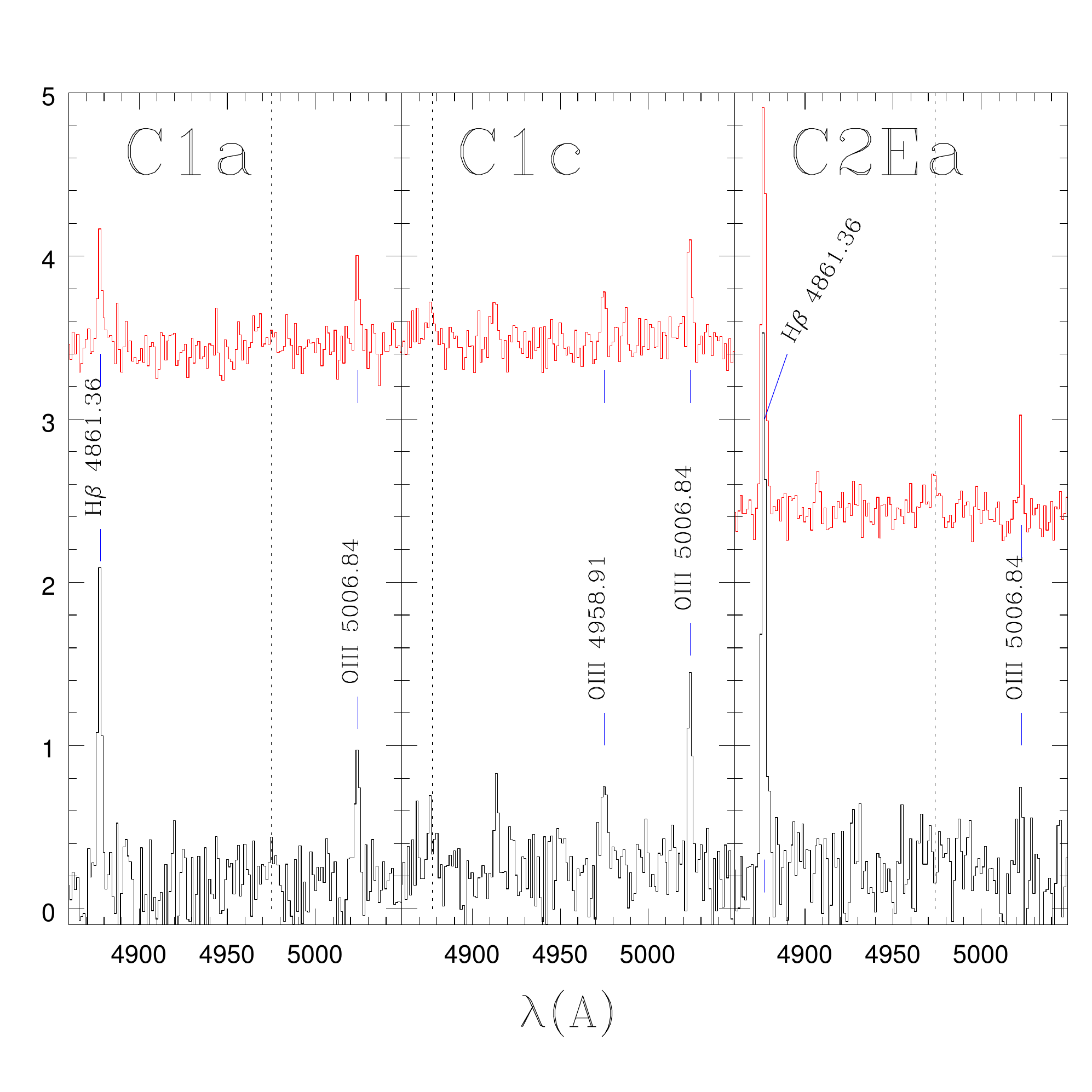}
 \caption{Detected and  undetected nebular line emission in the blue portions of the spectra for  3 regions of the Leo ring.  Apertures have radii of 2.4\arcsec  (black lines) and of 1.2\arcsec  (red lines). Emission lines detected at least in one aperture are labeled with the rest frame wavelengths and blue tick marks. Dotted lines  for undetected lines are placed at the expected wavelengths. Line intensity units along the y-axis are  10$^ {-17} $~erg~s$^ {-1} $~cm$^ {-2} $~A$^ {-1} $. Spectra have been arbitrarily shifted along the y-axes for display purposes.}
\label{spectra}
\end{figure}

\begin{table*}
\caption{Coordinates and integrated emission for gaussian fitted  lines in nebular regions.} 
\centering                                       
\resizebox{19cm}{!}{%
\begin{tabular}{c c c c c c c c c c  }           
\hline\hline 
 Source & RA & DEC &    H$\beta$  &  [OIII]4959 & [OIII]5007 &  H$\alpha$ &  [NII]6583 & [SII]6716/6731 & V$_{hel}$ \\  
 \hline\hline 
C1a&   10:47:47.9& 12:11:32.0    &5.19$\pm0.68$ & $<1.44 $        &  2.61$\pm$0.71         &  17.24 $\pm$0.51&  3.12$\pm$0.48  &  1.94/1.32$\pm$ 0.41  & 994$\pm$2  \\
C1b&   10:47:47.4& 12:11:27.7    & $<1.80  $          & $<1.71  $       &  $<1.71 $                     &    6.24 $\pm$0.45& 1.79$\pm$0.46   &   $<0.83$               & 1003$\pm$3  \\
C1c&   10:47:46.0& 12:11:08.6  &    $<1.56 $           & 1.89$\pm$0.69 & 3.83$\pm$0.71       &   2.17 $\pm$0.47&  $<0.76$               &   $<0.76$               &  1000$\pm$13   \\
C2Ea&  10:48:13.5& 12:02:24.3   & 10.1$\pm0.86$ &  $<1.52$            &  $<1.52$               & 35.47 $\pm$0.66 &  16.83$\pm$0.65 &   4.27/2.68$\pm$0.44 &  940$\pm$3   \\
C2Eb&  10:48:14.1& 12:02:32.5  &  $< 1.52 $         &  $< 1.52 $         &   $<1.52 $            &  4.32 $\pm$0.63&   2.34$\pm$0.50   &   $<1.33$               &937$\pm$21\\
 \hline\hline 
\end{tabular}
\label{lines}
}
\tablefoot{Emission line fluxes are measured in circular apertures with $R_{ap}$=2.4" and are given in units of 10$^{-17}$~erg~s$^{-1}$~cm$^{-2}$. Mean peak velocities V$_{hel}$ are in km~s$^{-1}$.}
\end{table*}

\subsection{The BPT diagram}

\begin{figure} 
\includegraphics [width=9 cm]{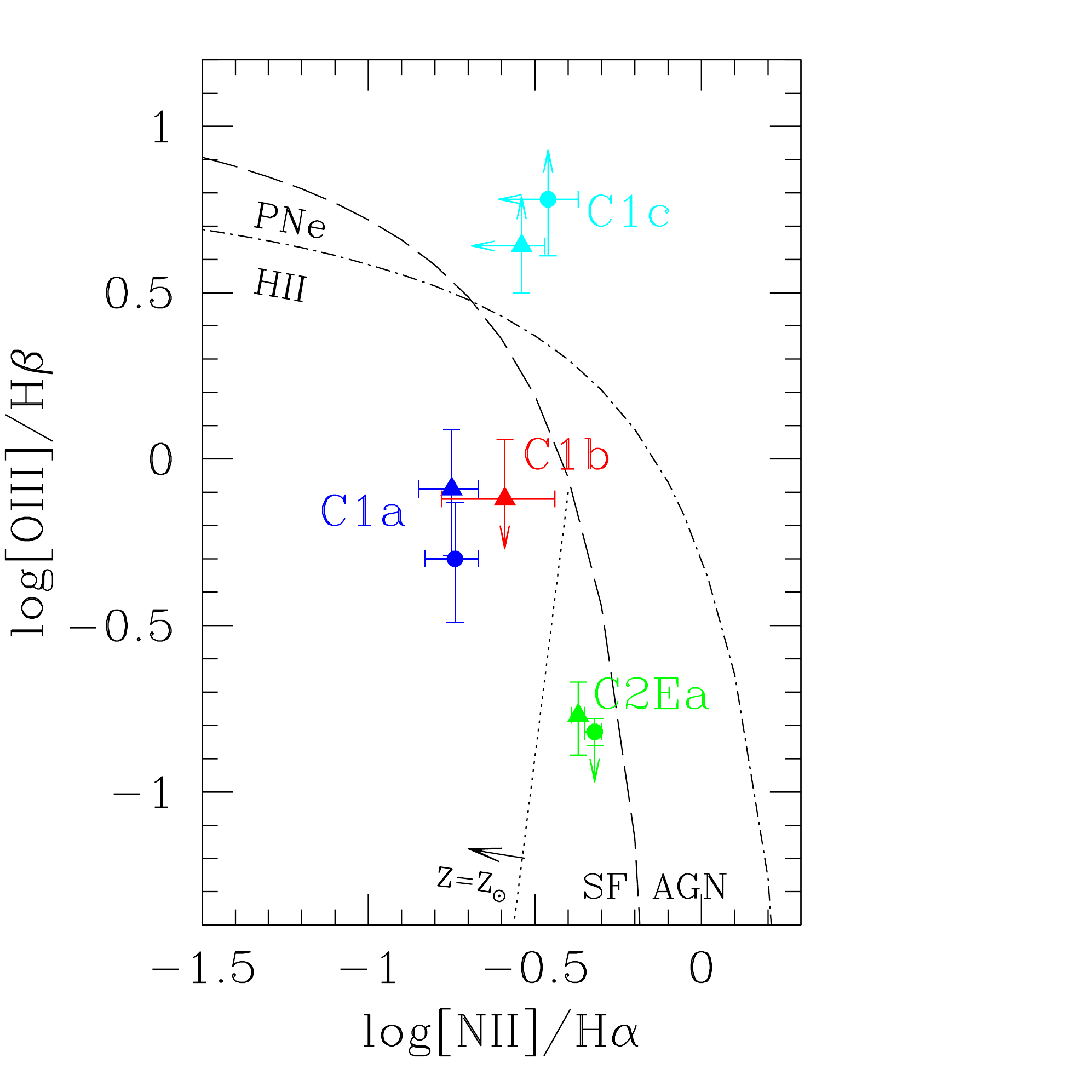}
\caption{The line ratios  [OIII]5007/H$\beta$ and  [NII]6586/H$\alpha$  are plotted for all nebular regions  in Table~1  for which  these ratios have been measured or limiting values can be inferred.  Different colors indicate different regions. Data for the largest aperture (radius 2.4\arcsec ) has been plotted with a filled circle and are listed in Table~2, filled triangles refer to 1.2\arcsec\  apertures.  For reference we also indicate the criteria proposed by \citet {2003MNRAS.346.1055K} for distinguishing between star forming galaxies and AGN (dashed line) and by \citet{2012ApJ...758..133S} to separate \ion{H}{II} regions  and PNe (dash-dotted line). All  \ion{H}{II} region evolutionary models of  \citet{2020MNRAS.496..339P} at Z=Z$_\odot$ fall to the left of the dotted line. The [NII]6586 line is  undetected in C1c and the [OIII]5007 line is undetected in  C1b.
}
\label{bpt}
\end{figure}

The diversity of the detected nebulae is marked in their spectra. In Figure~\ref{spectra} we show the blue side of the spectra relative to 3 nebulae: C2Ea has a strong H$\beta$ and a weak [OIII]5007 line while the opposite is true for C1c, where H$\beta$ is undetected but we detect also the [OIII]4959 line. The H$\beta$ and [OIII]5007 have comparable strengths for C1a. In Figure~\ref{bpt} we place 4 of the 5 nebular regions  on the [OIII]5007/H$\beta$ versus [NII]6586/H$\alpha$ plane, known as the BPT diagram (Baldwin et al. 1981). The line ratios  [OIII]/H$\beta$ and  [NII]/H$\alpha$  are plotted for all detected gaseous nebulae in the ring  for which these ratios have been measured or  their upper or lower limits can be determined, as indicated by the arrows.  Different colors indicate different regions  with the filled circle and triangles marking ratios for $R_{ap}$=2.4\arcsec\ and $R_{ap}$=1.2\arcsec\ respectively. For C1c, for which we have no  H$\beta$ flux, the [OIII]5007/H$\beta$ lower limit has been estimated inferring H$\beta$ from the unabsorbed  expected ratio H$\alpha$/H$\beta$=2.86. We indicate in Figure~\ref{bpt} the criteria proposed by \citet {2003MNRAS.346.1055K} for distinguishing between star forming galaxies and AGN  and by \citet{2012ApJ...758..133S} to separate \ion{H}{II} regions and PNe.  

As already stated in Paper I, the locations of C1a, C1b, and C2Ea  in the diagram are consistent with those of  \ion{H}{II} regions. As shown in Figure~3 they are close to the steeper descending part of the  \citet {2003MNRAS.346.1055K} line  where the  [OIII]/H$\beta$ ratio is low. In this region we expect \ion{H}{II} regions  metallicities close to or above solar and a  ionization parameter U which in dex  should be of order of $-3$ $-3.5$ \citep{2017ApJ...840...44B, 2017MNRAS.465.1384C}, compatible with our estimated ionizing fluxes and gas densities (see next Section). The location of C2Ea in the diagram requires higher than solar metallicities, in agreement with the metallicities inferred in Paper I and also with the results of recent simulations of evolving solar metallicity \ion{H}{II} regions in cold clouds \citep{2020MNRAS.496..339P} which populate the BTP diagram to the left of the dotted line in Figure~\ref{bpt}.  
 
\subsection{A candidate Planetary Nebula}

The location of C1c on the BPT diagram and other diagnostics suggest that we should consider this region as a candidate PNe rather than an \ion{H}{II} region, as outlined below in more detail.

\subsubsection{Position of C1c on diagnostic diagrams}

The nebula C1c lies above the \citet{2003MNRAS.346.1055K}  line for pure star forming galaxies. In this region [OIII]5007 lines are stronger than H$\alpha$ lines  and C1c is the only region where also the weaker  [OIII]4959 line has been detected. Following \citet{1981PASP...93....5B} a series of diagnostic diagrams has been developed to distinguish PNe from \ion{H}{II} regions \citep{2008MNRAS.384.1045K,2012ApJ...758..133S}. Unfortunately the H$\beta$ line is barely visible and we are forced to use its upper limit derived from H$\alpha$ line intensity for  Case B recombination  which gives log[OIII]/H$\beta$>0.64 and 0.78 for $R_{ap}$=1.2\arcsec\ and 2.4\arcsec\  respectively. We have applied only minimal foreground extinction A$_V$=0.074 \citep{1998ApJ...500..525S}; any additional extinction will increase the [OIII]/H$\beta$ ratio. No [SII] or [NII] lines are detected and the upper limits (at 3$\sigma$ level) imply log[NII]/H$\alpha<-0.46$  and log[SII]/H$\alpha<-0.46$ for the largest aperture.  As shown by Figure~\ref{bpt} C1c is within the BPT diagram region hosting mostly PNe rather than \ion{H}{II} regions \citep{2012ApJ...758..133S}.  

\citet{2020MNRAS.496..339P}  have recently computed very detailed models for the evolution of solar metallicity star forming region taking into account also the embedded phase, the shell emission while it expands, stellar winds and supernovae as well as and the possibility of multiple star formation events. This implies that collisional ionization due to shocks in the nebula and the star formation efficiency in the molecular cloud are also taken into account. An increase in  [OIII]/H$\beta$ ratio is predicted at very early times. The increase comes as the adiabatic expansion decreases the internal pressure and the density of the shell but this phase should not last very long. This might explain the region C1c  if the effectve line ratio [NII]/H$\alpha$ is a factor 3 or more below the limiting value, and [OIII]/H$\beta$ is of the order of the limiting value.  With the data presented in this paper and until higher sensitivity spectroscopic data is available, C1c better qualifies as a candidate PNe.  

PNe form the most luminous phase of evolution of their host stars with a non negligible fraction (as high as 10$\%$) of their luminosity emitted in the [OIII]5007 line. Given the line flux $F$[OIII] in units of erg~s$^{-1}$~cm$^{-2}$, one defines the apparent [OIII] magnitude as 

\begin{equation}
m_{5007}=-2.5 \log(F{\rm [OIII]})-13.74 
\end{equation}

Another  indication that C1c might be associated with a PN is given by the location of  C1c  in the diagnostic diagram  [OIII]5007/(H$\alpha$+[NII]) versus  $M_{5007}$ the absolute magnitude of [OIII]5007 line.  We find that  the nebula C1c is within the PN cone defined by \citet{2008ApJ...683..630H} where many other PNe lie. 

\subsubsection{The PNe luminosity function}

It is well established that the PNe luminosity function (PNLF), which gives the fraction of PNe in each [OIII] luminosity bin, has a steep cut-off at its bright end, as the absolute magnitude $M_{5007}$ approaches $-4.5$. This is almost invariant between different galaxies, with a slight dependence on metallicity and with the brightest PNe having  $M_{5007}^*\simeq-4$. Searches for PNe in galaxies of the Leo group have been carried out by \citet{1989ApJ...344..715C} and the PNe have been used as standard candles to determine their distances. The brightest PNe in NGC\,3384, a possible progenitor of the Leo ring, is $m_{5007}$=25.6 ($M_{5007}=-4.4$).  These bright PNe are likely descendant of stars more massive than 2~$M_\odot$ \citep{2004A&A...423..995M} or result from a binary evolution route \citep{2005ApJ...629..499C} although the identification of PNe progenitors is not yet settle \citep{2015ApJ...804L..25B}.  The evolutionary timescale for stars with mass between 2-5~$M_\odot$  is 0.1-1~Gyr while for a binary coalescence there is peak during the first 1~Gyr of the starburst event and then their number progressively reduces with time. For C1c, before extinction corrections, the apparent and absolute magnitudes are: 

\begin{equation}
m_{5007}=27.37^{+0.24}_{-0.20} \qquad  M_{5007}=-2.63^{+0.24}_{-0.20}
\end{equation}

The apparent magnitude is compatible with the negative result of the PNe survey by \citet{2003A&A...405..803C} which had 27.49~mag as completeness limit. The foreground extinction implies $A_V$ of 0.074~mag but the internal extinction is unknown although measurements suggests that $A_V<0.5$~mag are common \citep{2012ApJ...758..133S}.  Excluding extreme extinction values with $A_V>$1~mag which apply to PNe in galaxy disks with high gas column density along the line of sight,  the extinction corrected magnitude range for C1c is  $26.3<m_{5007}<$27.6. The candidate PNe is at least about one magnitude fainter that the PNLF bright cut-off value. The negative results of  \citet{2003A&A...405..803C} suggest that  this object might be the brightest PNe in the Leo ring with an intrinsic luminosity of the [OIII] line  $>$100~L$_\odot$. If the progenitor of this PN is a binary system or an isolated star with mass above 2~$M_\odot$ the evolution into a PN takes less than 1~Gyr and it is compatible with an in situ formation  in the case of  collisional origin of the Leo ring 1~Gyr ago.

\begin{figure} 
\includegraphics [width=9 cm]{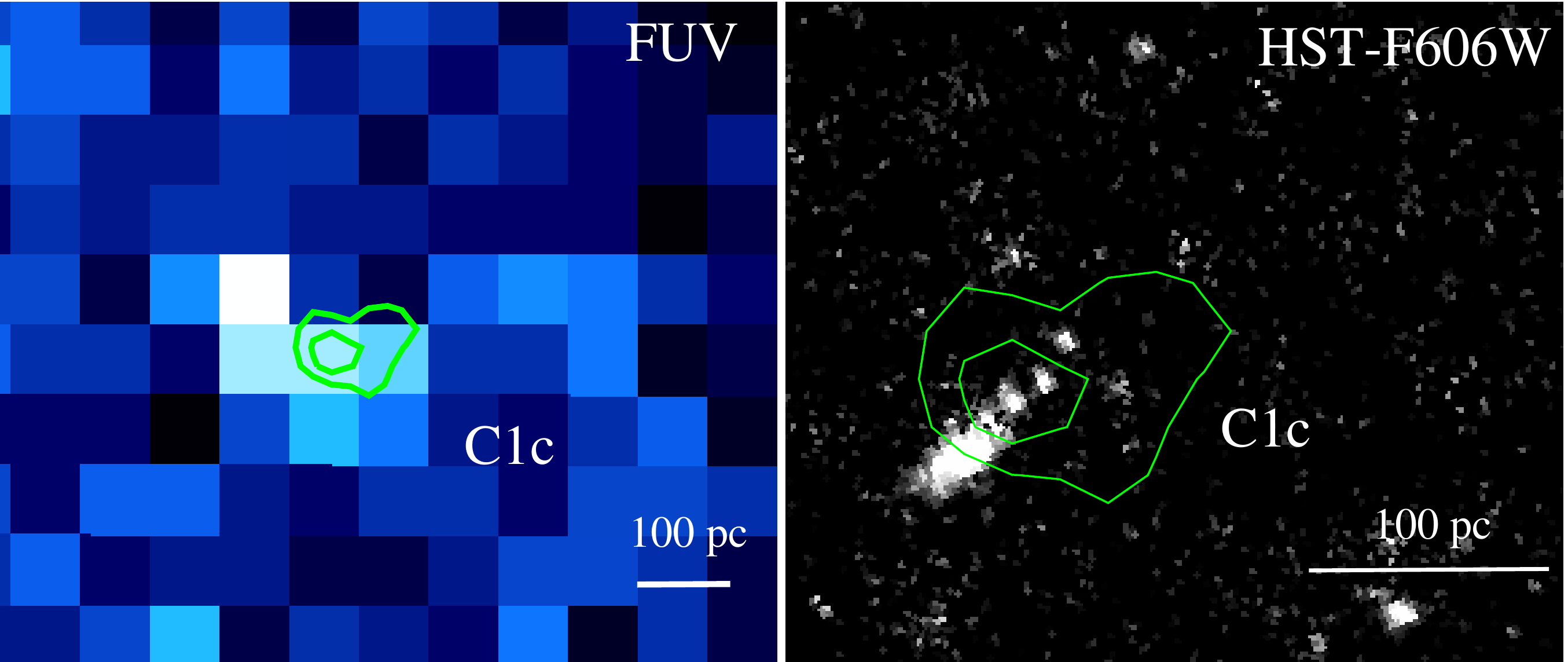}
\caption{Contours of the H$\alpha$ emission of the faintest emission nebula detected in Clump1, C1c, at 2.4 and 4$\times$10$^{-20}$~erg~s$^{-1}$~cm$^{-2}$ per pixel. The background images are the GALEX continuum FUV in the left panel and the HST-ACS-F606W optical image in the right panel.
}
\label{clump1c}
\end{figure}

The H$\alpha$ luminosity of C1c is consistent with the peak of H$\alpha$ PN luminosity function  measured in nearby galaxies such as M33 and the LMC \citep{2010PASA...27..149C}. In the HST images the H$\alpha$ emission of C1c, shown in Figure~\ref{clump1c},  overlaps with 3 unresolved optical sources which lie within a 1\arcsec\  region, next to a more extended source which might be a background galaxy. The optical counterpart of the PNe might be one of these unresolved knots in the HST image which have apparent red magnitude (F606W) of 26.6, 26.8, and 27.3$\pm0.2$, compatible with the bright end of the PNe luminosity distribution. In Figure~\ref{clump1c} we show also the very faint FUV emission which  might be associated with this source. Given the GALEX resolution, the presence  of spurious positive pixels, and the nearby background galaxy, we consider the faint FUV counterpart as "unlikely" association. 

\section{\ion{H}{II} regions  and their massive stellar population}

The presence of massive stars in the Leo ring implies a ionizing spectrum in the nebulae not much different than that of a galactic \ion{H}{II} region  \citep{2017ApJ...840...44B} and justify the computation of metallicities according to methods calibrated on bright \ion{H}{II} regions. In this Section we analyze stellar tracers across a variety of spatial scales:   from parsec scale point-like optical continuum emission, to HII regions traced by H$\alpha$, to  FUV emission across a few hundred parsec areas. We examine some characteristics of the massive stellar population powering the ionizing radiation such as their stellar spectral type, age and Initial Mass Function (IMF).   Massive stars are found in the field  but this may not represent their original environment since OB stars may be expelled via binary ejection events \citep{2017ApJ...834...94S}.  Using optical images and the intensity of H$\alpha$ emission, we attempt to investigate whether massive stars in the ring are part of a compact low mass stellar cluster at the center of the nebulae. We then determine mass and age of the sparse stellar population across wide areas  in  the nebula surroundings.

\subsection{Methods: sampling stellar tracers at different spatial scales}

{\it The optical view} - Archival HST data are available for the nebulae in Clump1 and these can be used to  image the stellar population powering them.
Given the angular resolution of the ACS camera (pixel size=0.05\arcsec\ =2.4~pc) we might distinguish stars in clusters with an intrinsic full width half maximum (FWHM) larger than a few parsecs. Compact low mass clusters hosting a bright outlier might be however indistinguishable from isolated massive stars, given extinction and photometric uncertainties. We use the F606W filter red magnitudes (VEGAMAG) of individual point-like sources and the F606W-F775W color to build a color magnitude diagrams (CMD) and compare stellar evolutionary tracks with the data.  We use PADOVA  isochrones \citep{2012MNRAS.427..127B} for stars of solar metallicity and no extinction. The position of the main sequence on this diagram is not much sensitive to metallicity. Given the  available red bands  additional extinction will not affect much the colors but the models move slightly to the right, towards bluer colors, for the modest amount of extinction measured in our nebular regions. At the same time magnitudes decrease with stars becoming brighter. 

{\it The H$\alpha$ view} - The observed H$\alpha$ luminosity can be used to compute the ionizing photon rate $Q_H$ in the nebulae for Case B recombination and solar metallicity. If the powering source is a single massive star we refer to main sequence models  listed in Table~1 of  \citet{2005A&A...436.1049M} to infer its stellar spectral type. We measure visual magnitudes to check if these massive stars might be hosted by compact stellar clusters localized at the center of the nebulae. The ionizing photon rate $Q_H$ from a stellar cluster is a function of cluster age. For clusters younger than 5~Myr  we consider $Q_H/L_{H\alpha}$=9.3$\pm0.4 \times 10^{11}$~s$^{-1}$/(erg~s$^{-1}$)  as given by  \citet{2017ApJ...840...44B}. This ratio is up to a 1.3 factor lower for clusters between 5 and 10~Myr of age. 

Because the most massive outlier have masses  $M<100$~$M_\odot$,  any associated  stellar cluster cannot have a fully populated IMF  up to 100~$M_\odot$. Stochasticity in the IMF implies that the mass of the most massive star is not strictly linked to the cluster mass \citep{2009A&A...495..479C,2011A&A...534A..96S}.  Given the limited photometric coverage and the photometric uncertainties, we do not attempt to run  stochastic models and consider stellar cluster  with an IMF fully populated up to  $M_{up}$. These models  have  the largest cluster mass compatible with the observed optical continuum and the H$\alpha$ flux  \citep{2009A&A...495..479C,2010MNRAS.401..275W,2011A&A...534A..96S}.  

\begin{figure} 
\vspace{-3cm}
\includegraphics [width=9 cm]{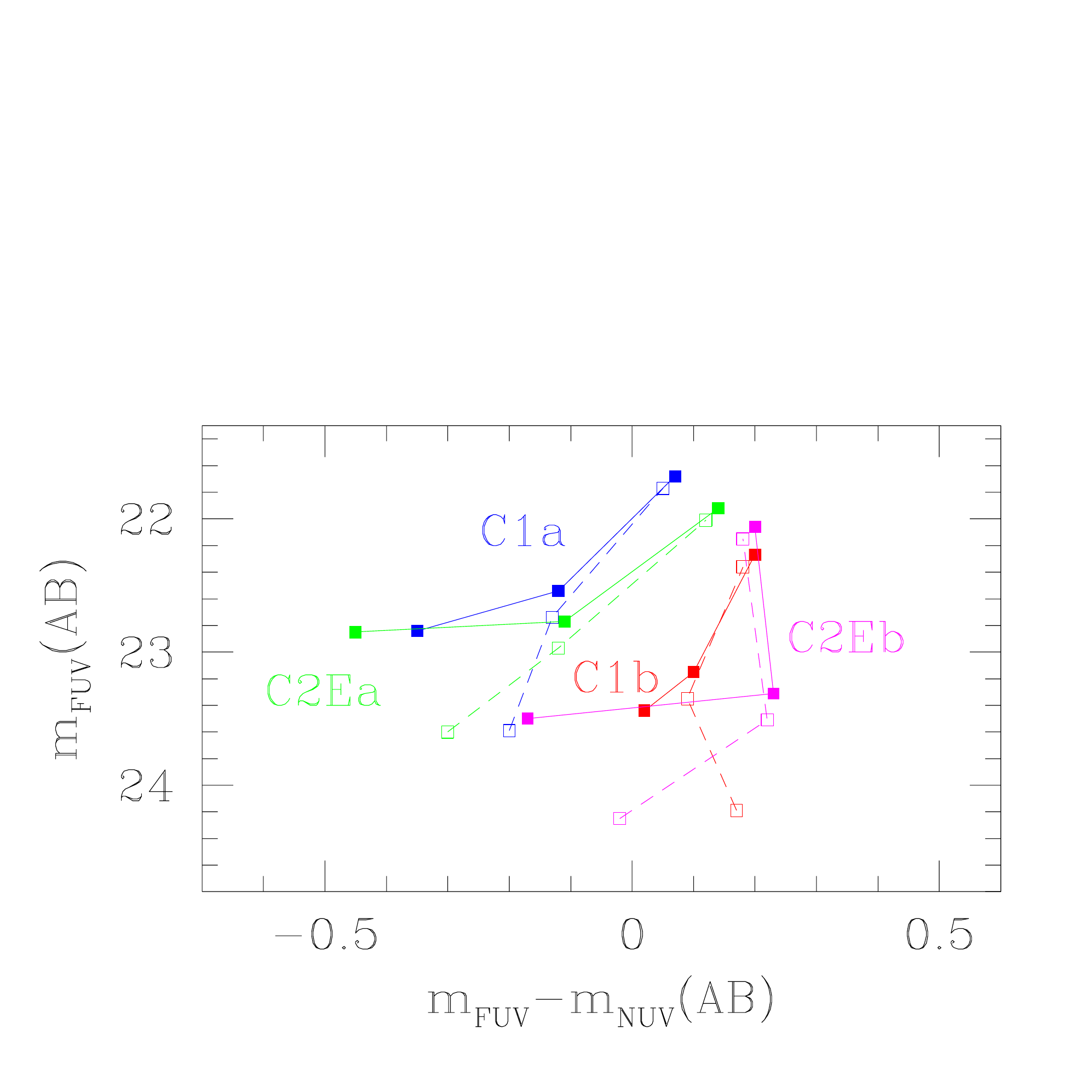}
\caption{Variations of the UV colors and FUV apparent magnitude as we vary the aperture size  with (continuous line) and  without (dashed line) aperture corrections. Along each line, from left to right, the apparent magnitude for the 5 nebular regions, color coded as in Figure~\ref{bpt}, decreases as we increase the aperture radius: from 2.4 to 3.8 to 7.5\arcsec  as marked by the square symbols.  } 
\label{uvcol}
\end{figure}

\begin{figure*} 
\includegraphics [width=18 cm]{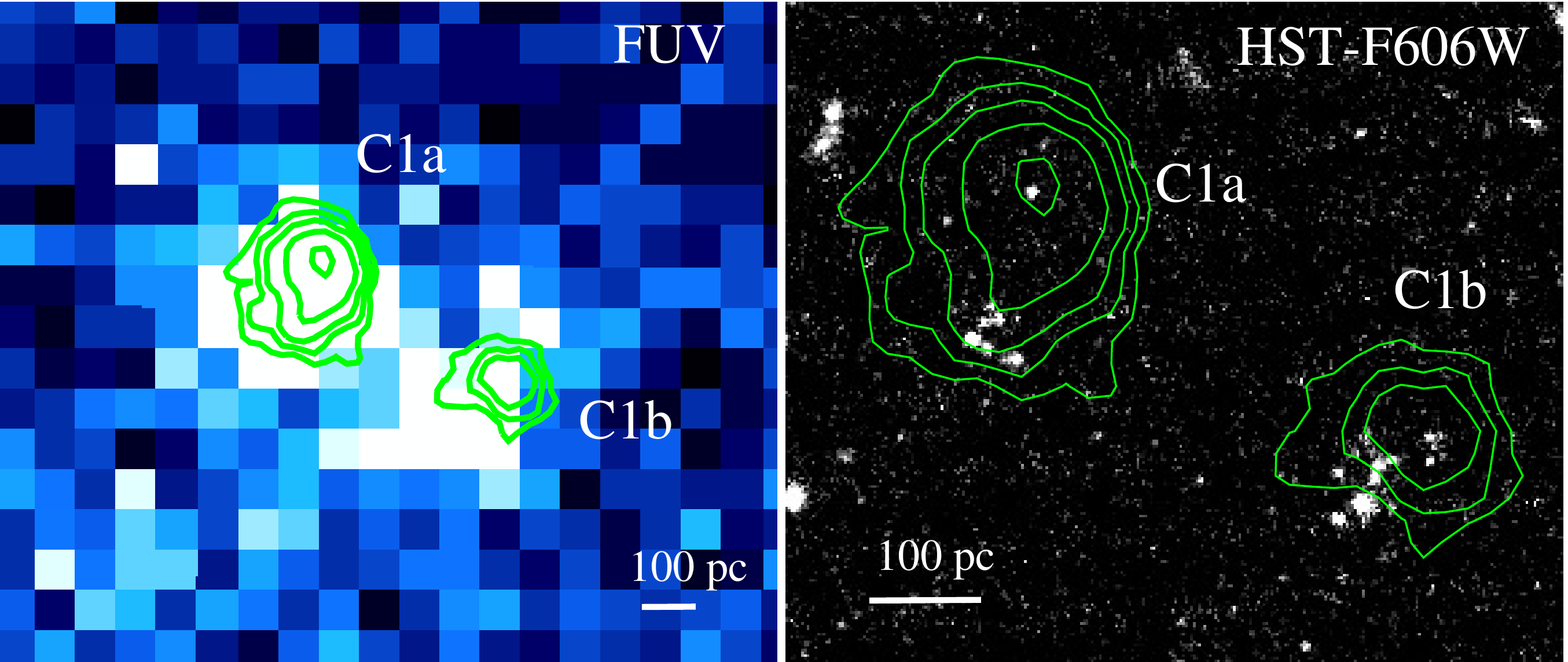}
\caption{ Contours of the H$\alpha$ emission of the two brightest \ion{H}{II} regions in Clump1, C1a and C1b, are overplotted in green to the GALEX-FUV continuum image in the left panel and to the HST-ACS-F606W optical image in the right panel. Contour levels are: 2.5,4,6,10,20$\times$10$^{-20}$~erg~s$^{-1}$~cm$^{-2}$ per pixel (0.2\arcsec ). The radius of the 10$\times 10^{-20}$~erg~s$^{-1}$~cm$^{-2}$ contour  level is about 70 and 50~pc for C1a and C1b respectively. The HST image shows that only part of the stellar population in the cloud is emitting ionizing photons powering the Str\"omgren spheres.}
\label{ha_hst}
\end{figure*}

{\it The UV view} - Photometry  in other bands at lower resolution, such as the UV, can help pinning down star formation properties across wide areas. The GALEX spatial resolution is 4.3\arcsec\ and 5.3\arcsec\  in FUV and NUV respectively. The UV colors are not uniform across Clump1 and Clump2E: they vary as we vary the aperture size in agreement with our finding that only part of optical and UV sources in the MUSE fields are associated with nebular emission, and hence with massive stars.  This might be due to some background contamination but also to age or mass spread in the stellar population.  Relevant aperture corrections should be applied for point-like sources \citep{2007ApJS..173..682M} when the aperture radius is $R_{ap}< 5$\arcsec . 
The UV counterparts of gaseous nebulae  however are not unresolved or point-like but rather diffuse and this limits the use of aperture corrections.  This can be seen by centring the aperture on the H$\alpha$ emission of the  \ion{H}{II} regions and computing the FUV and NUV AB magnitudes for $R_{ap}$=2.4,3.8,7.5\arcsec . The smallest aperture matches the  H$\alpha$ extent  of the brightest nebulae and the largest one has negligible aperture corrections.  Figure~\ref{uvcol}  shows the FUV magnitude and the FUV-NUV  color  before and after applying aperture corrections with dashed and continuous lines respectively. If sources were point-like we should have measured variations of order 0.7~mag in FUV between the smallest and the largest aperture, while we find more than one magnitude difference. These variations and color fluctuations indicate the presence of a diffuse  sparse stellar population next to the \ion{H}{II} regions.  

To minimise aperture corrections we analyze   the FUV emission in the rest of this paper by considering circular apertures  with $R_{ap}\ge 3.8$\arcsec (184~pc) , sampling areas larger  than the ionized nebulae. The FUV/H$\alpha$ luminosity ratio is computed for the four \ion{H}{II} regions using apertures with 3.8\arcsec  radius, after applying aperture and extinction corrections. This ratio will be very useful to determine mass and age of stellar bursts in a few hundred parsec regions around the nebulae. Photometric uncertainties are considered in addition to 50$\%$ uncertainties in extinction corrections for the FUV luminosity. 
We  compute the observed FUV luminosity by multiplying the spectral emission at 1538.6~\AA, the FUV band effective wavelength   \citep{2007ApJS..173..682M},  by the effective filter width (269~\AA). 
The UV background is computed over larger regions with no emitting sources.

\subsection{Optical photometry in Clump1}

The H$\alpha$ emission contours for C1a and C1b nebulae are shown in Figure~\ref{ha_hst}, and are overlaid on the GALEX-FUV image in the left panel, and on the HST-ACS/F606W image on the right panel. Both regions have a UV and optical counterparts. Unresolved point-like sources are clearly visible on the HST F606W image in the core of the nebular emission. The outermost H$\alpha$ contour of C1a bends toward a small group of stars, which might contribute to the ionization balance of the region. More than one source is visible toward the center of C1b, in addition to a  group of bright stars to the south east side of the H$\alpha$ peak.

\begin{figure*} 
\includegraphics [width=10. cm]{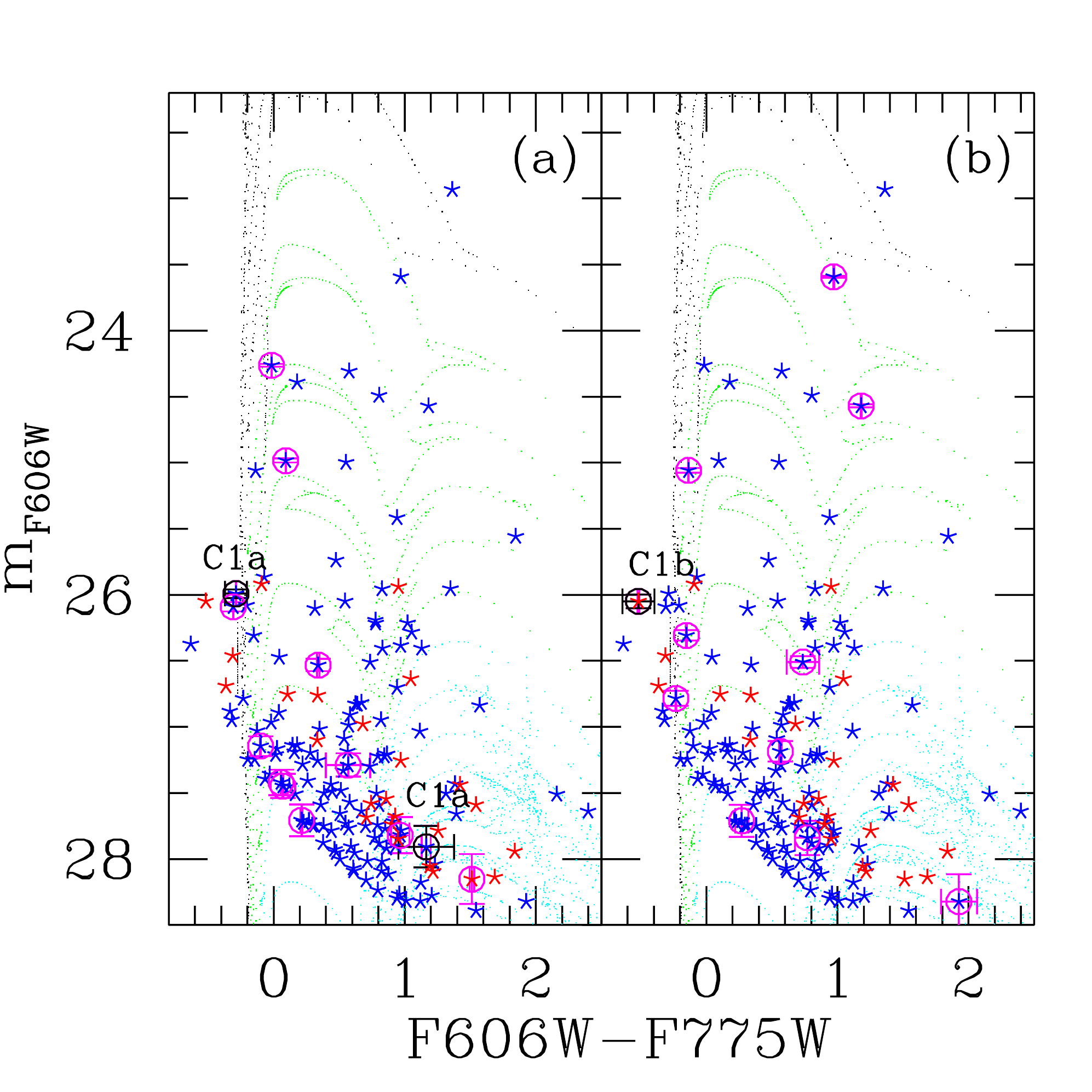}
\includegraphics [width=7 cm]{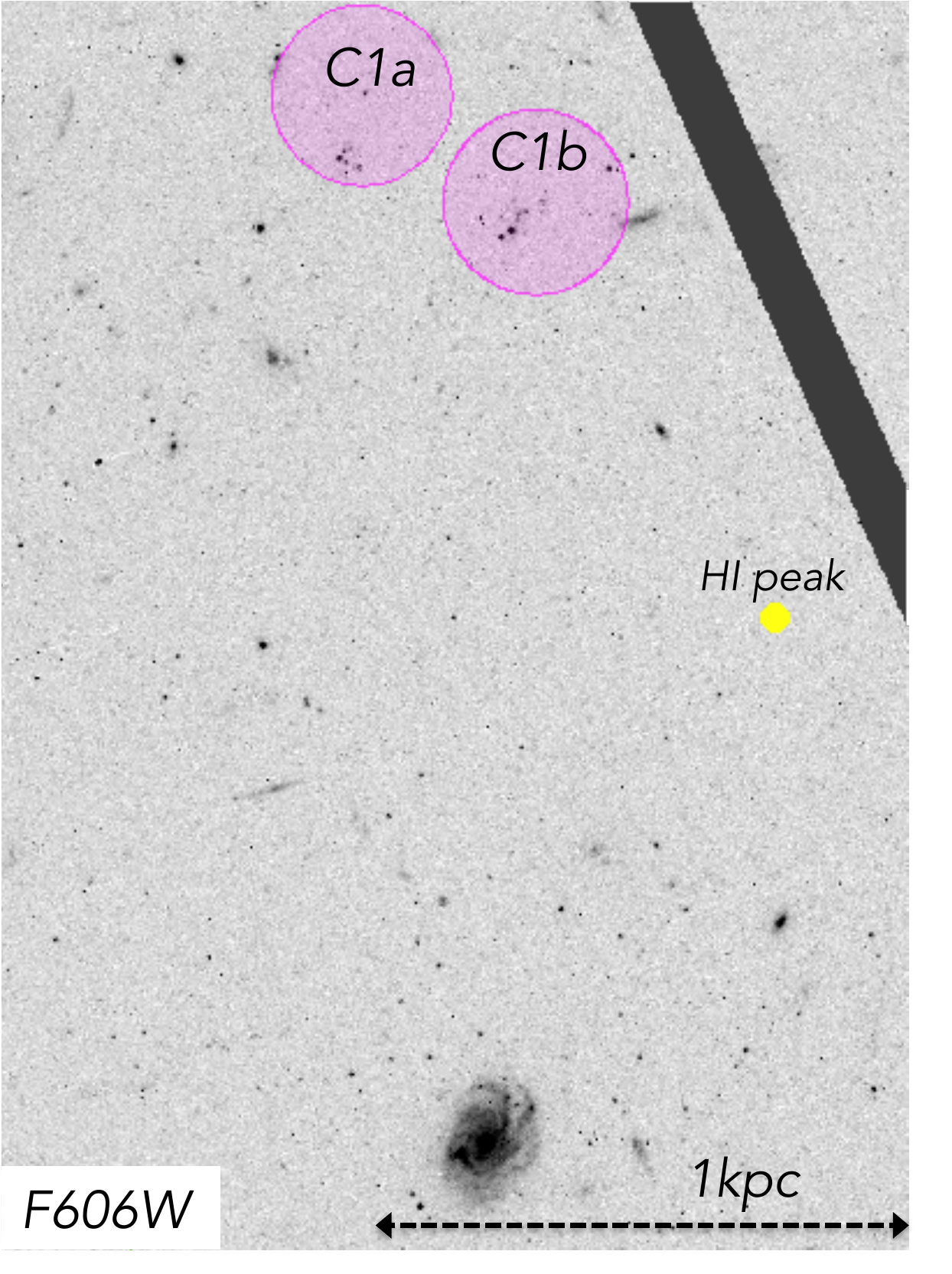}
\caption{Point-like sources within 1~kpc  of the center of the area hosting the two brightest nebulae in Clump1 are shown with asterisk symbols in the CMD (in VEGAMAGS) of panel (a) and (b).  The open magenta circles mark sources within 3.8\arcsec\ from the center of  C1a  in  panel (a) and of C1b in panel (b), with black color and labels used for sources at the center of the nebulae.  Blue asterisks are for objects  of type=1 (good stars). The dotted lines are the predicted evolutionary tracks for PADOVA isochrones for Z=0.0142, with black color for ages $\le 10$~Myr, green color for ages between 10 and 100~Myr and cyan color for ages $\ge 100$~Myr. No extinction corrections have been applied. To the right   the HST-F606W image shows the sparse population of stars in a region of Clump1. Magenta circles of 3.8\arcsec\ radii have been placed at the location of C1a and C1b, a filled yellow dot indicates the HI peak of Clump1. For reference the dashed line is 1~kpc in length.}
\label{cmd}
\end{figure*}

Isochrones have been plotted with dotted lines in Figure~\ref{cmd}$(a)$ and $(b)$. In the same Figure asterisks indicate sources extracted within a distance of 1~kpc from the position  RA=10:47:47.7 DEC=12:11:30 which is half way between the location of the centres of C1a and C1b. We have plotted  in blue sources which satisfy the following criteria in both F606W and F775W filter: {\it i)}the square of the sharpness  is $<0.2$ {\it ii)}the signal-to-noise is $>5$ 
{\it iii)}magnitudes are lower than the limiting magnitudes 
{\it iiv)}object type $=1$ i.e. good star candidates. Red asterisks satisfy the same criteria except {\it iv)} having object type $=4$ (nominally defined as ''no good star''). Black circles with labels highlight sources at the center of each  nebula, which might provide most of the ionizing photons. We use magenta colors to circle sources  which are within 3.8\arcsec\ distance of the center of  C1a  (in Figure~\ref{cmd}$(a)$)  and of  C1b (in Figure~\ref{cmd}$(b)$) i.e. within the magenta areas  overlaid to the HST F606W image in the right panel of Figure~\ref{cmd}.  

The isophotes for the object at the center of C1a  are slightly elongated and interpreted by DOLPHOT as two close stars. These are marked in blue because they satisfy all criteria for having a reliable point-like HST photometry.The $r$-band apparent magnitude of the brightest one is 25.99$\pm0.04$ with color F606W-F775W=$-0.29$. This translates into an extinction corrected absolute magnitude  $-4.4<M_V< -5$ (considering $V-R$=0 and  $A_V=0.4-1$~mag along the line of sight to the central star) suggesting a main sequence massive outlier with mass of the order of 30-40~$M_\odot$ (O7.5--O6-type).  The fainter companion is 2~mag fainter with F606W-F775W=1.1 and might be a 5~$M_\odot$ evolved star, 100~Myr old.  We cannot distinguish the two stars separately in the images and we cannot discard the possibility that wings in the profile of a single bright source elongate the isophotes. There is another star in magenta  close to main sequence at the 26th magnitude and very similar colors to the central star of C1a. This is a star in the south-east tail of the H$\alpha$ emission of C1a, where the \ion{H}{II} region contours bend (see Figure~\ref{ha_hst}).  Its associated H$\alpha$ emission seems however fainter than expected if this massive star  were on main sequence. The star might be embedded with some of the ionizing photons absorbed locally or  leaking out from the molecular cloud in the direction where the medium is more ionized  (the nearby massive star). Given the photometric uncertainties we cannot exclude that the  star is of lower mass, like 20~$M_\odot$, which is becoming brighter as it evolves off the main sequence.  
 
The two brightest evolved stars circled in magenta in the top part of the CMD are about 10-20~Myr old and have estimated masses of order 10-15~$M_\odot$:  each of them is located in the group of stars to the south east  of the C1a and C1b nebulae (within the H$\alpha$ contours shown in Figure~\ref{ha_hst}). The brightest central object of C1b has a magnitude similar to the central star of C1a but fails on the {\it iv)} requirement. Considering the  3$\sigma$ interval around the measured color, data are  still compatible with the star being on the main sequence  with a mass of order 20~$M_\odot$. The extinction corrected absolute magnitude is $-4.4<M_V<-4$ ($A_V=0.-0.4$~mag),  consistent with the identification of  an O8--9-type star. This is our best candidate for powering the C1b nebula, having reliable sharpness and magnitude uncertainties in each filter. Some fainter stars close to the center of C1b, but below the F775W limiting magnitude,  can contribute to the ionization balance only if local extinction is considerably higher than estimated for C1b or if they are in a binary system  \citep{2018MNRAS.477..904X}.

\subsection{Ionizing photons from massive outliers and the faint H$\alpha$ ring}

Given the extinction corrections and the extinction curve used in this paper (see Section~2) the total H$\alpha$ luminosity of the four \ion{H}{II} regions reads: $4\times 10^{36}$, $10^{36}$ $6\times 10^{36}$ and $0.9\times 10^{36}$~erg~s$^{-1}$ for C1a, C1b, C2Ea, C2Eb respectively.

The outermost H$\alpha$ contour of C1a in Figure~\ref{ha_hst} includes a region of about 300~pc  although most of the emission originates from the innermost 200~pc size region. For C1a we have $Q_H$ =4$\times10^{48}$~s$^{-1}$ suggesting an O7-type main sequence star with mass of the order 30~$M_\odot$ powering the nebula. This is in excellent agreement with the  visual magnitude of the central object from the HST photometry estimated in the previous subsection. Because of the additional direct absorption of ionizing photons by dust  or possible photon losses by leakage, the above inferred  $Q_H$ has to be considered a lower limit. If the rate of ionizing photons produced by the central object is twice larger than estimated from $L_{H\alpha}$ the central star can be as early as a O6.5 type. On the other hand, the most massive star in this area can be  slightly less massive if much more than a few percentage of the ionizing photons are emitted by the south-east group of stars where the contours bend. 

Given $L_{H\alpha}$ and $M_V$  for the central object of the C1a, we use instantaneous burst models of Starburst99 at solar metallicity with cut off mass $M_{up}\simeq$ 30-35~$M_\odot$ to infer a maximum stellar cluster mass $M_{cl}\simeq 300$~$M_\odot$ at the center of the nebula and an age of a few Myr.  A much less massive outlier or a much more massive stellar cluster would conflict with the data.  Given the uncertainties on extinction corrections, leakage of ionizing photons etc., both an isolated  massive star or a compact low mass cluster with a massive outlier are  consistent with the data  at the center of C1a. Similar considerations apply for C1b.

Since most of the C1a H$\alpha$ luminosity is coming from a 200~pc region, we can estimate  the volume density of this nebular region assuming a  homogeneous ionized Stromgren sphere with radius 100~pc  ionized by  an O7-type star. We find n$_H$=0.4~cm$^{-3}$,  at the lower extreme for typical \ion{H}{II} region densities, but  compatible with the average \ion{H}{I} gas volume densities estimated by \citet{1986AJ.....91...13S} for Clump1 (0.15~cm$^{-3}$). We can use this density to infer the ionization parameter in the nebula C1a defined as
 
 \begin{equation}
 U={Q_H\over 4\pi R_S^2 n_H c}
  \end{equation}
 
 where c is the speed of light. We have a ionization parameter with log U of order -3.5.

\begin{figure*} 
\includegraphics [width=18 cm]{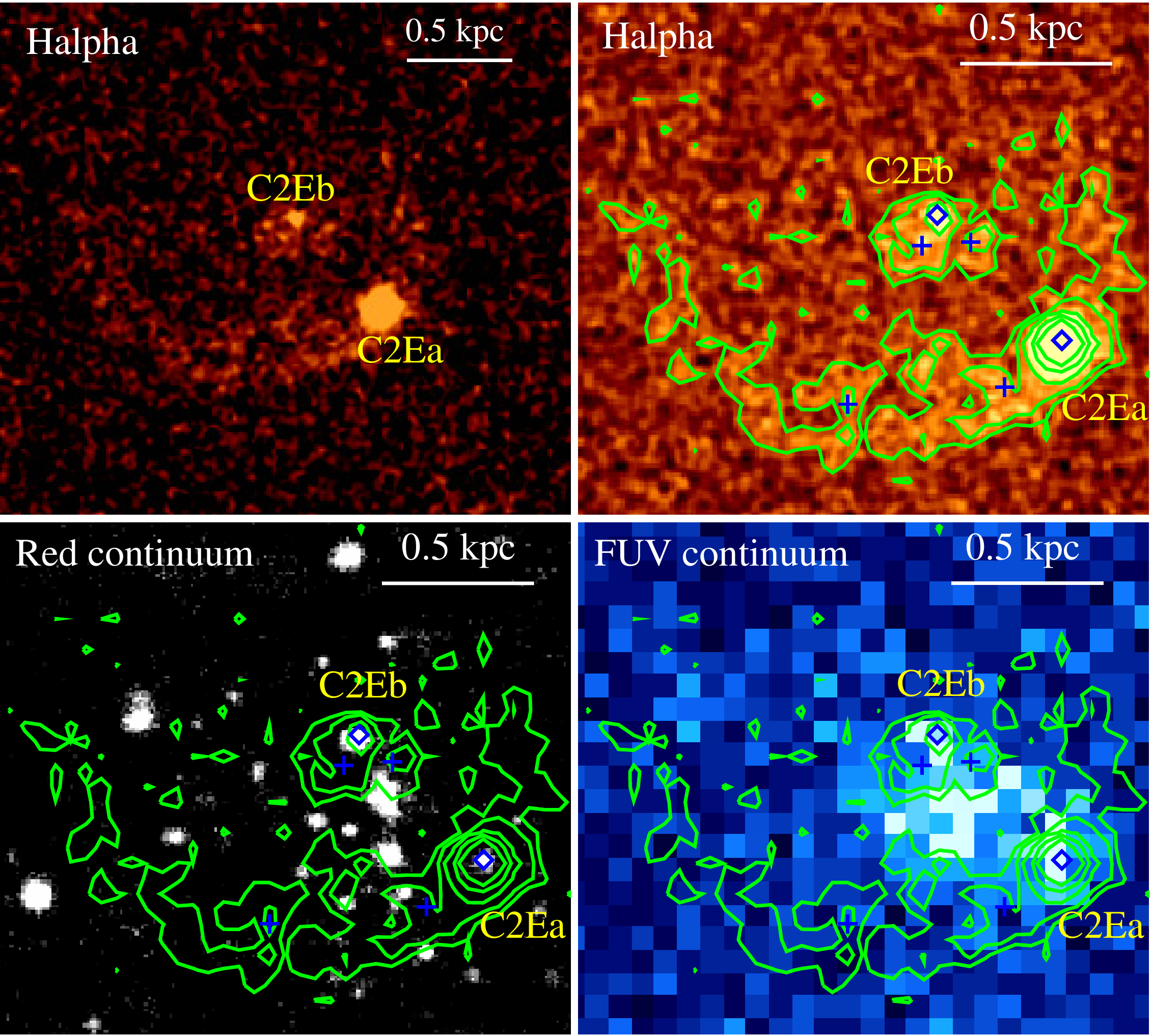}
\caption{The image of Clump2E in H$\alpha$ (log scale) is shown in the {\it upper-left} panel.The two brightest \ion{H}{II} regions, with more than one nebular line detected, are marked with blue diamonds in a zoom in image of  the H$\alpha$ emission in the {\it upper-right} panel (linear scale). The H$\alpha$ smoothed contour levels  at 1.2,2,4,10,20$\times$10$^{-20}$~erg~s$^{-1}$~cm$^{-2}$ per pixel shape   a partial ring of radius $\sim 0.6$~kpc. They are marked  also on the VLT red continuum image ({\it bottom-left} panel) and on the FUV-GALEX image ({\it bottom-right} panel). The blue crosses,  at the location of the H$\alpha$ sources  listed in Table~3, are some examples of the marginally detected  faint H$\alpha$ emitters with no optical counterpart.}
\label{fig_C2E}
\end{figure*}

Clump2E  hosts two nebular regions: C2Ea, the brightest \ion{H}{II} region detected by MUSE, and C2Eb.  They are shown in Figure~\ref{fig_C2E}. C2Ea is  brighter  in H$\alpha$  than C1a but slightly less extended. The corresponding rate of ionizing photons, 6$\times 10^{48}$~s$^{-1}$, is compatible with the presence of an O6.5 type star (35~$M_\odot$).  Considering possible direct absorption of ionizing photons by dust, the star powering this region can be as massive as a O6 type.  
In the bottom-left panel of Figure ~\ref{fig_C2E} we display the MUSE continuum image of Clump~2E where the weak optical counterpart  of C2Ea is visible and clearly fainter than the C2Eb optical counterpart. The continuum image of Clump2E reveals several sources  which might be older stellar clusters or associations,  undetected through nebular line emission, or background galaxies. In the bottom right panel a FUV counterpart of the nebulae is visible in addition to emission linked to other continuum sources. 

In the {\it top-left} panel of Figure~\ref{fig_C2E} we show the 1~arcmin$^2$ continuum subtracted  H$\alpha$ map of Clump2E observed by MUSE using a log scale and  spatial smoothing. Some faint H$\alpha$ emitters can be seen.  These lie, together with C2Ea, along a partial ring with projected radius of about 13\arcsec  (0.6~kpc) centered on C2Eb.  The  H$\alpha$ peaks  and integrated fluxes  are listed for some of them in Table~2. The contours of the H$\alpha$ intensity are displayed on the central 33$\times$33~arcsec$^2$ region of the MUSE continuum image and of the FUV-GALEX image in Figure~\ref{fig_C2E}. 
No other lines and no visible optical counterparts are detected at the location of the faint H$\alpha$ emitters. They can be embedded \ion{H}{II} regions  where star formation is taking place, or warm knots of  an extended shock front that  propagated in the \ion{H}{I} clump and stimulated new star formation events. The shock might have been originated from a recent tidal interaction, by multiple supernova explosions or by the expansion of the second bright \ion{H}{II} region in Clump2E, at the ring center, powered by a very massive star that already ended its life.

\begin{table}
\caption{Coordinates and H$\alpha$ emission for some of the faint  nebulae in Clump2E } 
\centering                                       
 \begin{tabular}{ccccc}           
\hline\hline 
Source& RA& Dec& H$\alpha$-peak & H$\alpha$-flux \\   
 \hline\hline 
C2Ec&  10:48:14.16& 12:02:30.5& 0.15$\pm$0.03 &0.44$\pm0.13$   \\
C2Ed&  10:48:13.94& 12:02:30.7& 0.09$\pm$0.02 &0.19$\pm0.06$  \\
C2Ee&  10:48:13.77& 12:02:21.3& 0.06$\pm$0.02 &0.21$\pm0.08$   \\
C2Ef &  10:48:14.48& 12:02:20.1& 0.10$\pm$0.03 &0.21$\pm0.09$   \\
 \hline\hline 
\end{tabular}
\label{faint}
\tablefoot{Peaks and fluxes are in units  of 10$^{-17}$~erg~s$^{-1}$~cm$^{-2}$~\AA$^{-1}$ and 10$^{-17}$~erg~s$^{-1}$~cm$^{-2}$ respectively.}
 \end{table}

\subsection{Stellar burst models } 

\begin{figure*} 
\center
\includegraphics [width=13 cm]{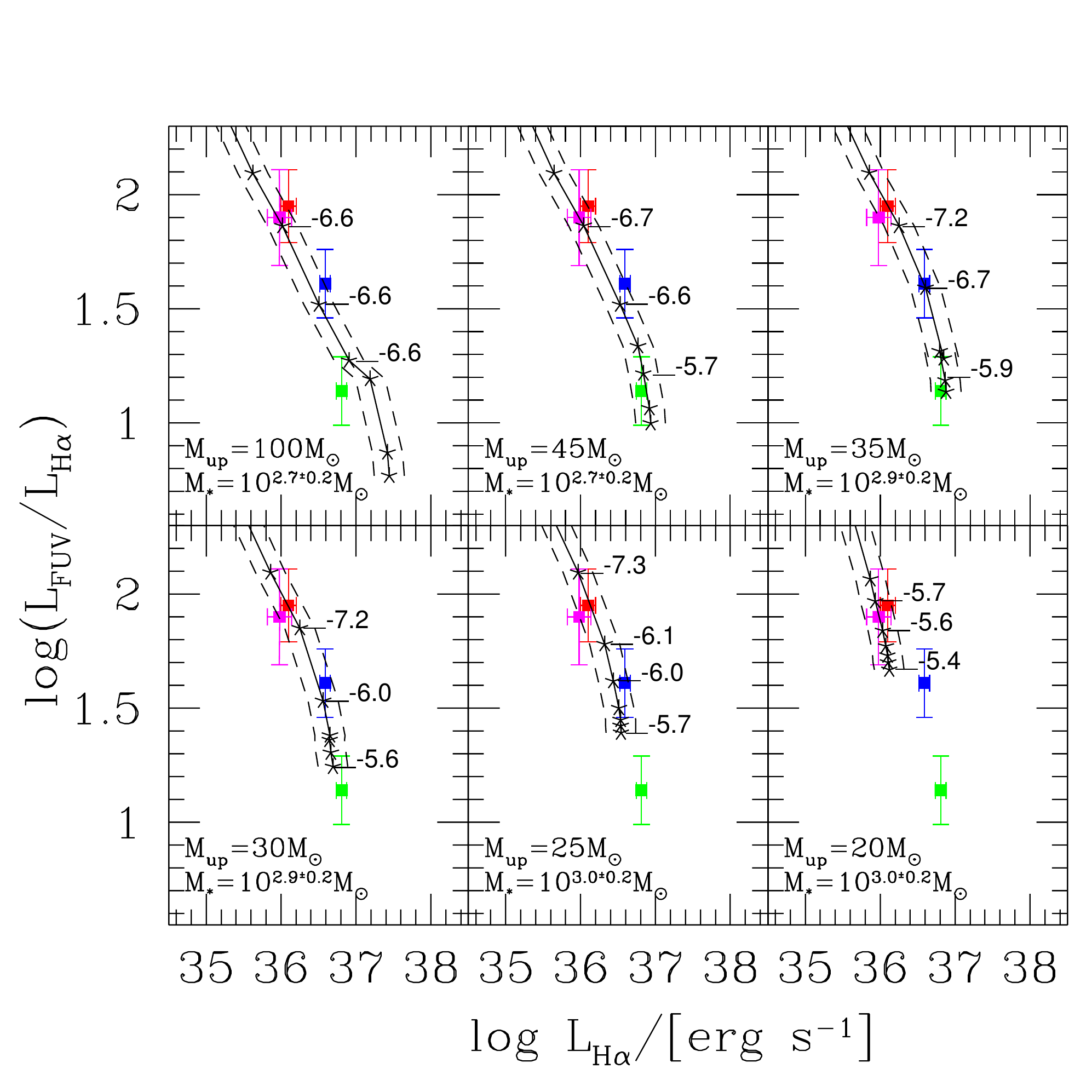}
\caption{Time evolution of extinction corrected total H$\alpha$ luminosity of the regions as a function of FUV/H$\alpha$ luminosity ratio.  The continuous line in each panel refer to a solar metallicity burst model with stellar mass and IMF upper end mass cut-off as indicated by $M_{*}$ and $M_{up}$. Time increases from bottom to top along the curves with asterisks marking time steps of 1~Myr from 1 to 7~Myr. The $M_{*}$ value in each panel has been chosen to best match the evolutionary model with the data,  dashed lines indicate  model predictions  for 0.2 dex variations in $M_{*}$.  Some theoretical values of absolute visual magnitudes during cluster evolution are printed to the right of  the horizontal tick marks indicating the time step. The extinction corrected value of FUV/H$\alpha$ luminosity ratio for the 4 \ion{H}{II} regions, color coded as in Figure~\ref{bpt}, is plotted with filled squares and is measured in circular apertures with a 3.8\arcsec  radius. }
\label{burst6}
\end{figure*}

If we  consider a coeval birth of stars  in the nebulae and their close surroundings,  the FUV-NUV color increases with time especially after 100~Myr. However at this time  the cluster nebular  emission is negligible. We are interested in younger clusters, which emit a non negligible amount of photons blueward of 912~\AA.  At these early times UV color variations with age are very modest and a very accurate UV photometry is needed to estimate ages. Moreover, a complication arises  because for low star formation densities the IMF is not fully populated up to its upper mass end and in the first 10 Myr UV colors depend on the massive stellar population. The  lack of massive stars might increase the FUV/H$\alpha$ ratio which can be interpreted as an older age. 
To break the age-$M_{up}$ degeneracy we plot in Figure~\ref{burst6} time variations of FUV/H$\alpha$ luminosity ratio as a function of H$\alpha$ luminosity. In the chosen apertures nebular recombination lines are well detected and  apertures for the two \ion{H}{II} regions in each Clump do not overlap. 

A few Myr after the birth of a stellar cluster  the H$\alpha$ emission declines with time faster than the FUV flux, due to the stronger dependence of the H$\alpha$ flux from the massive stellar population. As time increases a cluster  moves upward and to the left in the diagrams of Figure~\ref{burst6}.  
The continuous line in each panel refer to a Starburst99  \citep{1999ApJS..123....3L} instantaneous burst model with solar metallicity, stellar mass and IMF upper end mass cut-off as indicated by $M_{*}$ and $M_{up}$ labels respectively. Time increases from bottom to top along the curves with asterisks marking time steps of 1~Myr from 1 to 7~Myr. The $M_{*}$ value in each panel has been chosen to match the evolutionary models with the data and refers to the continuous lines;  dashed lines are model predictions for 0.2 dex variations around $M_{*}$.  The mean age of the populations seems well constrained and independent on the IMF upper mass cut-off:  C2Ea is younger than  C1a which is 5 Myr old, while C1b and C2Eb are the oldest regions (6-7~Myr old).  An increase/decrease of the stellar mass moves the evolutionary models to the right/left. It is then possible to find a value of $M_*$ that best fits the data for each star forming region except for C2Ea which requires $M_{up}\ge30~M_\odot$. It is remarkable however that all the data can follow a unique burst model with similar $M_{up}$ and stellar masses. We might be detecting similar  populations but at different post burst times.

The additional constrain which should be considered is the visual magnitude associated to each star forming region and to each model. Theoretical values of absolute visual magnitudes during cluster evolution are printed to the right of horizontal tick marks at some time step of the models. They refer to the continuous line and vary by $\pm0.5$ between the two dashed lines i.e. as we vary the cluster mass by 0.2~dex. The extinction corrected visual magnitudes of the optical counterparts of the two regions in Clump1 have been estimated using the HST-DOLPHOT extracted sources (which satisfy the conditions $i)$-to-$iii)$) and are $M_V$=-6.9$\pm0.3$ and -7.4$\pm0.2$  for C1a and C1b respectively. Integrated photometry over the region gives somewhat higher values (by 0.8 mag) due to contamination from some diffuse background source. Visual magnitudes include the contribution of evolved stars which are not strictly coeval with the burst powering the nebular emission and should be considered as upper limits. For Clump2E we have used the VLT/MUSE continuum image which give $M_V$= -6.0$\pm0.2$ and -7.3$\pm0.2$  for C2Ea and C2Eb respectively. Visual and FUV magnitudes over circular apertures with 3.8\arcsec\  radius confirm then the presence of a coarse stellar population in the area  of the \ion{H}{II} regions, as shown in Figure~\ref{ha_hst} and Figure~\ref{cmd}, with magnitudes consistent with the predictions of a local bursts. A close inspection of Figure~\ref{burst6} shows that individual burst models with $M_{up}$ in the range 45--30~$M_\odot$ provide good fit to the whole dataset. They predict stellar masses across a 0.1~kpc$^2$ area of order 500-1000~$M_\odot$ and stellar mass densities of order 0.005--0.01~$M_\odot$~pc$^{-2}$. The optical colors $B-V$ of the models for the matched ages and masses are in agreement with that measures by \citet{2014ApJ...791...38W}. 

\section{The star formation rate} 

Massive stars in the clumps  indicate unambiguously that star formation is taking place in situ in the ring during the last 5~Myr.  The star formation rate (SFR) across the ring can be estimated   using  the extended GALEX-FUV emission maps which gives mean values of the SFR over the last 100-300~Myr.  Possible incomplete sampling of the IMF has a major impact on estimates of the SFR based on H$\alpha$ line rather than on that based on the continuum FUV. To infer the SFR we run several models with Starburst99  with a continuous star formation rate at solar metallicity but choosing different mass cut-offs at the upper mass end. Table~\ref{conv} shows the $L_{\nu,FUV}/L_{H\alpha}$ ratio and the conversion factors $C_{FUV}$, $C_{H\alpha}$ between the luminosity at the GALEX FUV effective wavelength  (in erg~s$^{-1}$~\AA$^{-1}$) or the H$\alpha$ luminosity (in erg~s$^{-1}$), and the SFR in $M_\odot$~yr$^{-1}$ according to the following equations:

\begin{equation}
{{\dot M_{FUV}\over [M_\odot~{\rm yr}^{-1}]}} = C_{FUV} {L_{\nu,FUV}\over [{\rm erg~s}^{-1}~\AA^{-1}]}
\end{equation}

\begin{equation}
{{\dot M_{H\alpha}}\over [M_\odot~{\rm yr}^{-1}]} = C_{H\alpha}{L_{H\alpha}\over [{\rm erg~s}^{-1}]} 
\end{equation}

Table~\ref{conv} shows the expected $L_{\nu,FUV}/L_{H\alpha}$ ratio  for different IMF upper mass cut offs $M_{up}$. Results are shown for a Chabrier IMF and for times longer than 100~Myr past the start of  star formation. Stochastic sampling of the IMF implies that massive stars might form at a random time in the upper portion of the IMF which might not be fully sampled.  In this case the  SFR has a spread  due to stochasticity and the mean SFR, as for a truncated fully populated IMF,  is also underestimated if one uses the standard conversion factor  with $M_{up}=100~M_\odot$. Deviations from the true SFR are larger if H$\alpha$ emission is used as SFR  indicator  \citep{2014MNRAS.444.3275D}.  The bias is smaller if the SFR is traced by FUV emission, but it is not very large also when H$\alpha$ emission is used  if SFR$\ge 3\times10^{-4}$~M$_\odot$~yr$^{-1}$. Specific peak and mean values  of the SFR distribution for  stochastic sampling of the IMF depend on the choice of the prior. We remark that for a given cluster mass, the stochastic sampled IMF with the occasional formation of massive outliers predicts a large scatter in the FUV-to-H$\alpha$ luminosity ratios than the IMF model with a fully sampled IMF up to a fixed mass cut-off \citep{2009A&A...495..479C}. The  FUV-to-H$\alpha$ peak ratios are however similar to what Table~\ref{conv} shows.

\begin{table}
\caption{ Star formation rate coefficients } 
\centering                                       
\resizebox{8.5cm}{!}{%
\begin{tabular}{cccc}           
\hline\hline 
$M_{up}$ & $C_{FUV}$ &$C_{H\alpha}$ & $L_{\nu,FUV}/L_{H\alpha}$\\   
$M_{\odot}$ & $M_\odot$~yr$^{-1}$/[ erg~s$^{-1}$~\AA$^{-1}$] &$M_\odot$~yr$^{-1}$/[ erg~s$^{-1}$]&[erg~s$^{-1}$~\AA$^{-1}$]/[erg~s$^{-1}$]\\   
 \hline\hline 
100&   9.1 10$^{-41}$  &  5.7 10$^{-42}$    &    1.3  \\
45 &   1.3 10$^{-40}$   &  1.2 10$^{-41}$    &    1.5  \\
35 &   1.4 10$^{-40}$   &  1.9 10$^{-41}$    &    1.7  \\
30 &  1.5 10$^{-40}$    &   2.8 10$^{-41}$   &    1.8  \\
25 &   1.6 10$^{-40}$   &   4.4 10$^{-41}$   &    2.0  \\
20 &   1.7 10$^{-40}$   &   1.0 10$^{-40}$   &    2.3  \\
\hline\hline 
\end{tabular}
}
\label{conv} 
\tablefoot{Coefficients and the expected luminosity ratio $L_{\nu,FUV}/L_{H\alpha}$ are computed for a Chabrier IMF with upper end mass cutoff at $M_{up}$}
\end{table}

\subsection{The star formation rate in nebular regions}

Continuous star formation models can be used to estimate the  star formation rate density  in  the MUSE fields where  H$\alpha$ and UV emission have been detected. A continuous star formation model traced by FUV emission refers to star formation over  at least 100~Myr.  Over such long timescales we cannot sample small areas which experience local bursts.  Star formation does not happen persistently in a given location but it  propagates in nearby regions due to local feedback and to its a stochastic character related to large scale gas instabilities \citep{2015ApJ...814L..30E,2018MNRAS.478.3793D}. Molecular clouds form where gas is compressed, have their lifecycle over 10-20~Myr and disperse back into the ISM as soon as the newborn massive stars evolve off the main sequence \citep{2017A&A...601A.146C}. For this reason the location of H$\alpha$ peaks shifts spatially from the molecular gas (pre-burst) and UV (post-burst) peaks. To sample properly star formation we use circular apertures with radius of 7.5\arcsec  (364~pc) centered on the H$\alpha$ peaks and correct for extinction. Additional absorption of  ionizing photons by dust grains implies a somewhat higher SFR than quoted here. 
We calibrate the continuous star formation rate with the observed ratio $L_{\nu,FUV}/L_{H\alpha}$ i.e. we evaluate $M_{up}$ and $\Sigma_{SFR}$ in each region using the prescription given in Table~\ref{conv}. We shall use the simulation of \citet{2014MNRAS.444.3275D} for a flat prior with $\sigma=0.25$ to evaluate the uncertainties on the SFR due to stochasticity.

The star formation rate densities $\Sigma_{SFR}$ over a 0.42~kpc$^2$ area of  the star forming regions detected in H$\alpha$ are shown in column 8 of Table~\ref{sfr}. The luminosity ratio $L_{\nu,FUV}/L_{H\alpha}$ used to determined $M_{up}$ and  $\Sigma_{SFR}$ is displayed in Table~\ref{sfr}. The star formation rate density is nearly uniform and  reads 3$^{+7}_{-2}\times 10^{-4}$~$M_{\odot}$~yr$^{-1}$~kpc$^{-2}$.  We underline the excellent agreement with the estimate of \citet{2009Natur.457..990T} for  Clump2E. For Clump1 our  mean estimate  is a factor 2 higher. The availability of H$\alpha$ emission and the release of the assumption of a fully populated  IMF  allows us to have a more accurate mean conversion factor between the observed emission and the rate of star formation. With the observed star formation rate density over 100~Myr we expect $B-V$=0.05 and $\mu_B$=27.2~mag~arcsec$^{-2}$. The region is brighter than optical survey limiting values  but we underline that these surveys are for large scale features and   small optically brighter clumps cannot be excluded. This expected brightness is coincident with that estimated for C1a from HST data in the previous Section.

\subsection{Star formation across the ring and its relation with the \ion{H}{I} gas density}

\begin{figure} 
\includegraphics [width=9 cm]{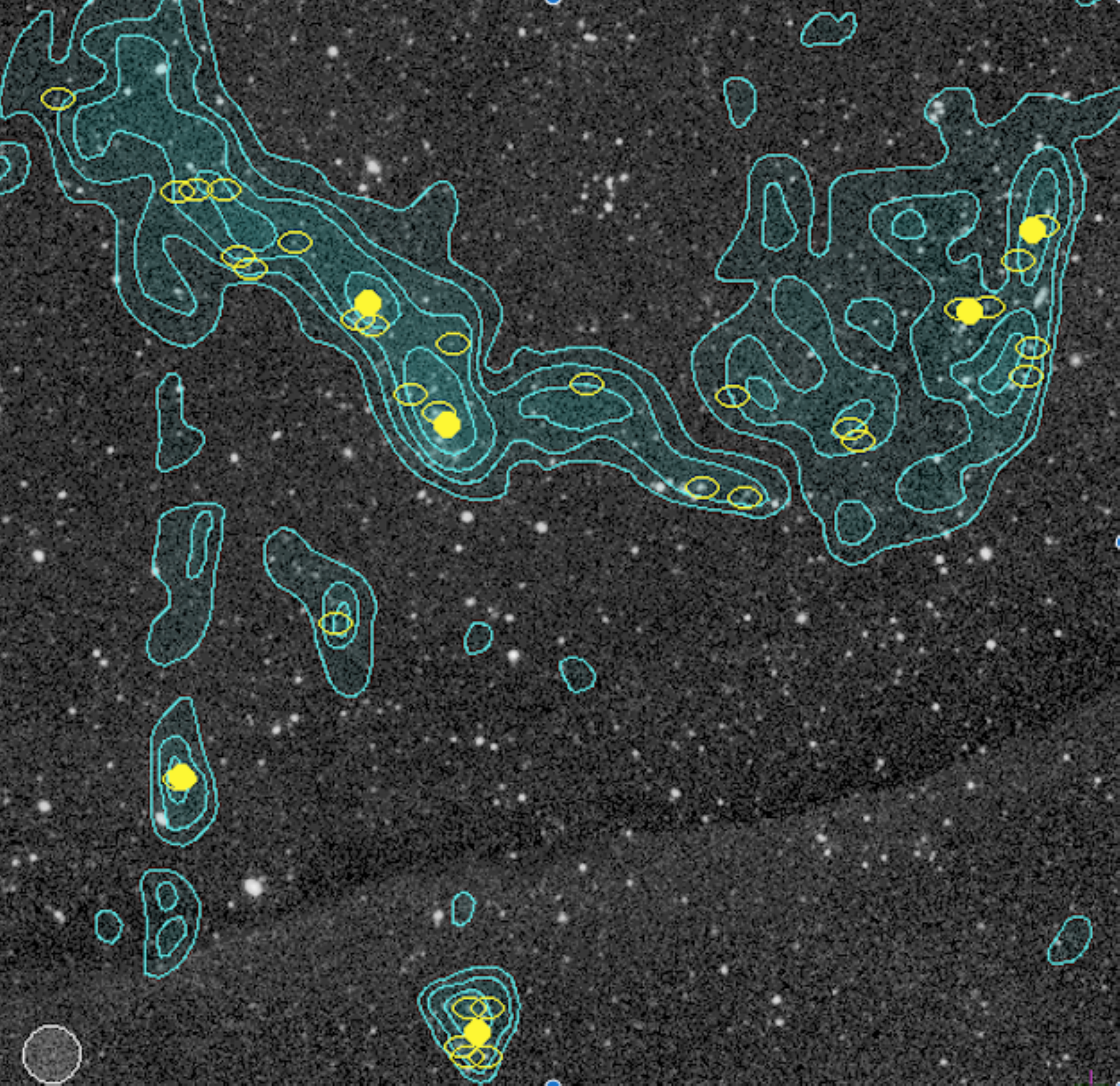}
\caption{Location of possible star forming sites  (yellow ovals) identified in the FUV-GALEX map of  the Leo ring (background image). Green contours are relative to 21-cm \ion{H}{I} emission mapped by \citet{1986AJ.....91...13S}. Filled yellow circles indicate the location of \ion{H}{I} peaks with associated FUV emission estimated using a circular aperture of 45\arcsec\  in radius, indicated by a gray circle at the bottom left corner.  
 }
\label{fuvsrc}
\end{figure}

We use FUV continuum to probe star formation across  large areas of the Leo ring. The GALEX maps inspected by \citet{2009Natur.457..990T} were less sensitive than the latest GALEX data release we are using and fainter star forming clumps can now be detected. Contamination from background objects is a relevant issue in this case. At the location of the FUV peaks where MUSE detected nebular emission, we noticed that no optical counterparts are visible. This is due to the lack of extended massive young star cluster in the ring whose light would be detectable also at a coarse spatial resolution. The HST data underlines in fact the presence of a sparse population of  stars, stellar association or compact low mass clusters. We have visually inspected the SDSS optical images for  FUV sources within 1\arcmin\ radius of \ion{H}{I} peaks in the cloud. To the \ion{H}{I} peak list provided by \citet{1986AJ.....91...13S} we added the \ion{H}{I} peaks present in the southern part of the ring, towards M\,96. We selected  FUV sources with no optical counterparts in the SDSS image as the most likely star forming sites in the ring. Faint background galaxies may still contaminate this sample but the FUV source proximity to \ion{H}{I} peaks sets these as the best candidate star forming regions. In Figure~\ref{fuvsrc} the cross symbols show the location of these sources on the FUV-GALEX map and \ion{H}{I} VLA contours. The spatial extent of each star forming region is between 10\arcsec\ and 30\arcsec\ (0.5-1.5~kpc). The southernmost clump with a few FUV peaks is BST1047+1156 where \citet{2018ApJ...863L...7M} found diffuse starlight  and UV emission.

Using circular apertures which match the UV size of the star forming regions we estimate a total FUV luminosity of these sources $L_{FUV}$=2.8$\times10^{39}$~erg~s$^{-1}$. This luminosity is computed by multiplying the observed $L_{\nu,FUV}$ (in erg~s$^{-1}$~\AA$^{-1}$)  by the effective FUV band width (269~\AA) for each UV clump, and then summing them up. 
Using this measure we infer a global SFR of order 1-2$\times10^{-3}$~$M_\odot$~yr$^{-1}$ following Table~\ref{conv}. For an average $A_V$=0.2 i.e. $A_{FUV}$=0.5~mag, the global SFR can be as high  as 3$^{+4}_{-2}\times10^{-3}$~$M_\odot$~yr$^{-1}$. The majority of these UV sources lie close or within the 10$^{20}$~cm$^{-2}$ \ion{H}{I} column density contour level which encloses a surface area of about  250~kpc$^2$. We can use this area to estimate the average SFR density in the denser part of the ring which  reads $\Sigma_{SFR}$=0.8$^{+2}_{-0.6}\times 10^{-5}$~$M_\odot$~yr$^{-1}$~kpc$^{-2}$.  A similar  FUV luminosity  and global SFR density is recovered if we sum the fluxes recovered in circular apertures centered  at the \ion{H}{I} peak locations. If this average SFR applies over 500~Myr and considering 30$\%$ mass loss  we infer an average stellar surface density of 0.003~$M_\odot$~pc$^{-2}$ and an apparent brightness $\mu_B\simeq$ of 31.6~mag~arcsec$^{-2}$ with $B-V$=0.1. This is consistent with the actual limiting values of large scale optical surveys of the Leo ring and with a planetary nebulae as bright as the candidate C1c discussed in Section 3. 

\begin{figure} 
\includegraphics [width=9 cm]{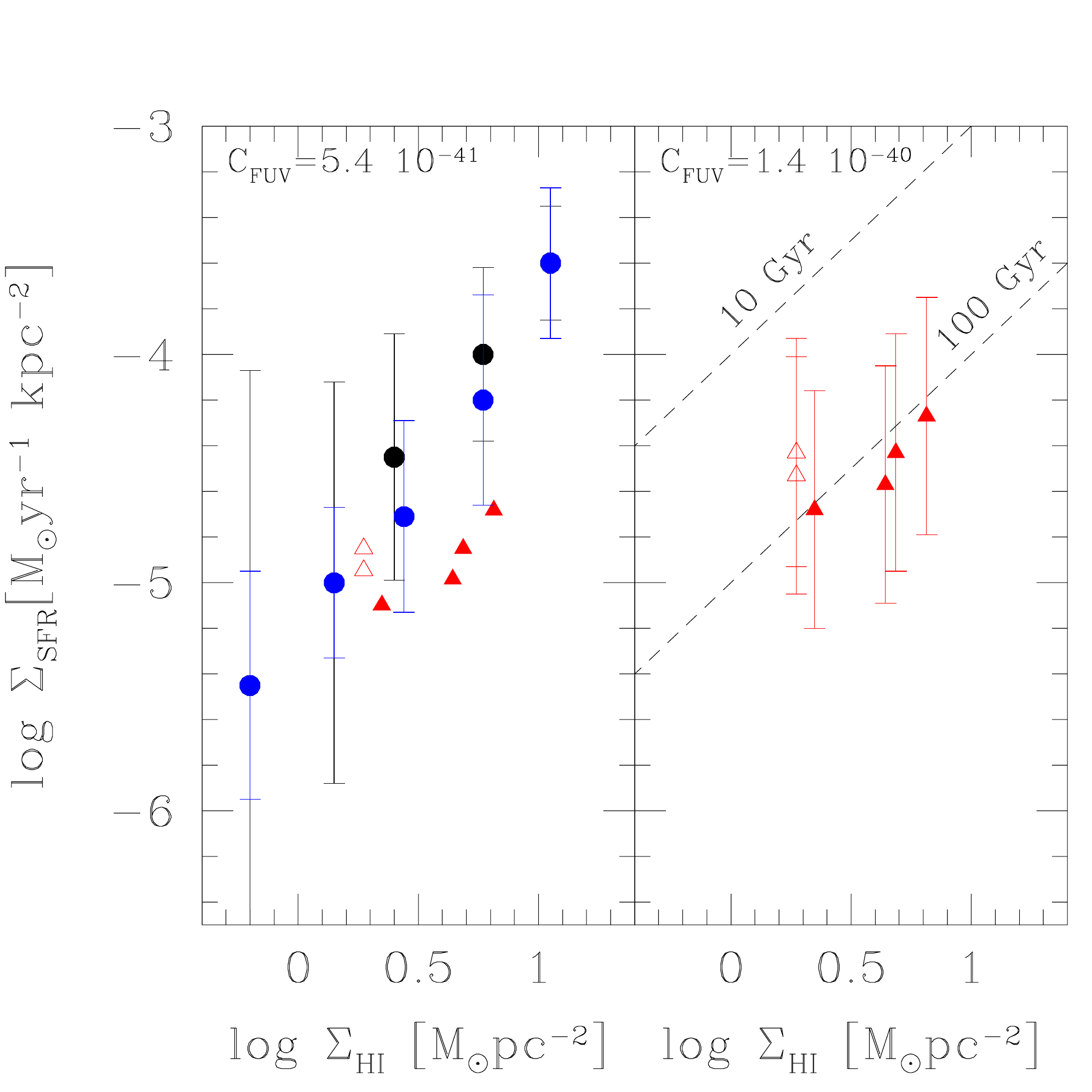}
\caption{ In the left panel FUV based estimates of the star formation rate densities $\Sigma_{SFR}$ and  \ion{H}{I} gas mass surface densities $\Sigma_{HI}$ are shown with red triangles for the 6 \ion{H}{I} peaks in the Leo ring with non negligible FUV emission. The open triangles indicate \ion{H}{I} peak data not listed by \citet{1986AJ.....91...13S} relative to cloudlets towards M96. Filled dots trace the relation for the median values of the large database  on outer disks of spiral galaxies (black color) and on dwarf galaxies (blue color)  obtained by \citet{2010AJ....140.1194B} and their dispersion. For this panel we use the same star formation rate coefficient as in  \citet{2010AJ....140.1194B}. In the right panel we show the star formation rate densities for the same 6 \ion{H}{I} peaks in the ring but computed with the conversion coefficient relative to Starburst99 continuous star formation models with $M_{up}$=35~$M_\odot$  and dispersion due to IMF stochastic sampling. The coefficients $C_{FUV}$ are in $M_\odot$~yr$^{-1}$/[ erg~s$^{-1}$~\AA$^{-1}$]. For reference we show the lines relative to constant depletion times of 10 and 100 Gyrs.  
 }
\label{ks}
\end{figure}

By employing a FUV photometric aperture with radius 45\arcsec, after subtracting the emission of UV clumps with optical counterpart, only 6 of the 25 \ion{H}{I} peaks show FUV emission above the noise, 4 in the main body of the ring and 2 in the tail towards M\,96 as shown by Figure~\ref{fuvsrc}. The brightest FUV emission is associated with  Clump1 where we also have a low  FUV-NUV color index, indicative of very recent star formation. Clump2E and Clump2 share similar colors in such large apertures,  and a slightly lower UV light intensity  than Clump1.  The total star formation rate is 3$^{+4}_{-2}\times10^{-3}$~$M_\odot$~yr$^{-1}$, similar to that estimated from individual  source counts. Being the aperture size twice as large as the VLA \ion{H}{I} beam size (FWHM=45\arcsec ), we can check the relation between the \ion{H}{I} gas mass  density and star formation rate density $\Sigma_{SFR}$ traced by the FUV emission for the 6  \ion{H}{I} peak regions. This relation, known also as the Kennicutt-Schmidt relation \citep{2012ARA&A..50..531K} is well established for the total gas mass surface density or for the molecular hydrogen mass density but  the relation between $\Sigma_{SFR}$ and the atomic gas density $\Sigma_{HI}$ has a large scatter.  However, several works on dwarf galaxies and outer disks of spiral galaxies, where the \ion{H}{I} column density is moderate and molecular hydrogen is undetected through its CO tracer, have outlined a  region in the   $\Sigma_{SFR}$--$\Sigma_{HI}$ diagram where the data lie and which we indicate in the left panel of Figure~\ref{ks} following  \citet{2010AJ....140.1194B}. We compute the \ion{H}{I} mass surface density  using the VLA data in Table~II of \citet{1986AJ.....91...13S}.   For the 2 southernmost peaks coincident with the cloudlets towards M\,96, which are not included in the list of  \citet{1986AJ.....91...13S}, we estimate gas masses of order 9.5$\times10^6$~$M_\odot$ over an extension of 5~kpc$^{2}$ from the \ion{H}{I} maps (i.e. a total \ion{H}{I} surface density $\Sigma_{HI}$ of 1.9~$M_\odot$~pc$^{-2}$). For this plot we use $C_{FUV}$= 5.4$\times 10^{-41}$ $M_\odot$~yr$^{-1}$/[ erg~s$^{-1}$~A$^{-1}$] to be consistent with the conversion factor used by \citet{2010AJ....140.1194B}. This is about a factor 2 lower than what Table~\ref{conv} gives for $M_{up}=100~M_\odot$ because it is based on a calibration derived by \citet{2007ApJS..173..267S} using specific population synthesis model fits to multiband photometric data. We correct for extinction (A$_{FUV}$=0.5~mag) although the dataset shown by  \citet{2010AJ....140.1194B} has not been corrected for extinction (expected to be low). Uncertainties in extinction estimates for these apertures are hard to quantify and dominates over photometric uncertainties.

The 6 candidate star forming regions in the Leo ring have gas and star formation rate densities compatible to those measured in dwarf galaxies and outer disks of spirals, although on the low side of the distribution. In the right panel of Figure~\ref{ks} we compute the star formation density according to the Starburst99 models for a continuous star formation rate at solar metallicity and $M_{up}$=35~$M_\odot$.   The uncertainties shown in the right panel are from the IMF stochastic sampled distribution with a flat prior of \citet{2014MNRAS.444.3275D}. Gas depletion times are of order of 100~Gyrs and ensure a long and lasting life to the ring. With MUSE observations presented in this paper we confirm very recent star formation in two of these regions and \citet{2018ApJ...863L...7M} inferred the presence of star formation on 100-200~Myr timescale in a third region towards M\,96, included in our list. Of the remaining 3 regions, one was partially covered by one MUSE field but  no nebular line emission has been detected (Clump2), and other two are on the west side of the ring main body. For the future we have already planned observations for detecting possible molecular gas  emission associated with the confirmed star forming regions.

\subsection{A summary Table}

Table~\ref{sfr} summaries our results on massive stars, stellar groups and star formation in the nebular regions. The name of the nebula and its total H$\alpha$ luminosity are in the first two columns. Follows the estimated spectral type of the central star mostly responsible of the ionization. Using the FUV-to-H$\alpha$ luminosity ratios, observed in a few hundreds pc regions and shown in column (4), we modeled the local burst and conclude that all the \ion{H}{II} regions are in a few hundred parsec size regions hosting 500-1000~$M_\odot$ of stars with similar IMF upper mass end. Estimated burst ages, quoted in column (6)  suggest only  2-4~Myr age difference between the bright and the faint \ion{H}{II} region in each clump. We fit the FUV-to-H$\alpha$ luminosity ratios relative to more extended regions around the nebulae, as listed in column (7), by using a continuum star formation model over at least 100~Myr  in these more extended regions. The average mass of the most massive star according to continuum star formation models in these regions is in column (8).  We recover a rather similar star formation rate density in the Clump1 and Clump2E: 3$\times10^{-4}$~$M_\odot$~yr$^{-1}$~kpc$^{-1}$ (column 9).  For  larger regions, of a few kpc in size, the rates can be read in the last column of Table~\ref{sfr}. 

\begin{table*}
\caption{ Massive stars and star formation rates  in the Leo ring}
\centering                                       
 \begin{tabular}{c c c c c c c c c}           
\hline\hline 
Source&  log $L_{H\alpha}$ & central & $L_{\nu,FUV}/L_{H\alpha}$ &  age     & $L_{\nu,FUV}/L_{H\alpha}$  &  $M_{up}$ & log $\Sigma_{SFR}$  & log $\Sigma_{SFR}$ \\  
             &  erg~s$^{-1}$      & stellar   & ${[{\rm erg~s}^{-1}~\AA^{-1}]\over [{\rm erg~s}^{-1}]}$ &  Myr   & ${[{\rm erg~s}^{-1}~\AA^{-1}]\over [{\rm erg~s}^{-1}]}$ & $M_\odot$  & $M_{\odot}$~yr$^{-1}$~kpc$^{-2}$  &$M_{\odot}$~yr$^{-1}$~kpc$^{-2}$  \\
             & total     &    type  &    $R_{ap}$=184~pc   &burst        &  $R_{ap}$=364~pc      & cont.  &  $R_{ap}$=364~pc   &  $R_{ap}$=2.2~kpc\\   
 \hline\hline 
C1a&    36.58  & O7      &1.6  & 5   &   1.8  &  30 &  -3.4  &  -4.3 \\ 
C1b&    36.06  & O8.5   &1.9  & 7   &   2.3  &  20 &  -3.5  &  -4.3 \\  
C2Ea&  36.79  & O6.5  &1.2  & 3   &   1.5  &  45 &  -3.5  &   -4.5 \\  
C2Eb&  35.97  & O9     &1.9  & 7   &   2.3  &  20 &  -3.5  &   -4.5 \\  
 \hline\hline 
\end{tabular}
\label{sfr}
 \end{table*}

The low star formation rate  in a metal and gas rich environment, underlines that the amount of metal pollution is not a key ingredient for the formation of stars in low density regions. Clearly the ability of the gas to condense and fragment is much more related to the growth of  local instabilities and perturbations which  drive the gas into a self gravity dominated regime  and provide self shielding to important coolant such as molecules \citep{2004ApJ...609..667S, 2019A&A...622A.171C}. These conditions are only occasionally fulfilled in outer disks due to their shallow potential, lower column density and stellar feedback. Since the early work of \citet{1989ApJ...344..685K} it became evident however that a low level of star formation and  faint HII regions can be found beyond the radius where H$\alpha$ drops  \citep{2005ApJ...619L..79T,2007ApJS..173..538T, 2005ApJ...627L..29G,2007ApJ...661..115G}. Here metallicity gradients flatten to values  below solar  but metal abundances are still high enough to require some mixing with metals produced elsewhere in the inner disk, or  accretion of metal rich gas  as galaxies evolve.  This and our results on the Leo ring underline that metallicity is less relevant than other physical conditions for determining the star formation rate density. The shorter gas depletion time in outer disks compared to the Leo ring, implies that  rotating outer disks next to the deep potential  wells of bright inner disks, develop more easily overdensities and cold gas filaments \citep{2012ApJ...757...64B} which later fragment to form stars.

\section{Summary and conclusions}

Spectroscopic observations using MUSE operating at the Very Large Telescope of ESO have revealed the presence of ionized gas and metal lines in gas overdensities with \ion{H}{I} masses of order 10$^7$~$M_\odot$ in the giant Leo ring. Two of the three MUSE fields centered on \ion{H}{I} peak locations have  nebular lines which have allowed us to determine reliable chemical abundances close to or above solar, and a low extinction. 
These chemical abundances coupled to an undetected diffuse stellar counterpart of the ring has been used in Paper I to constrain the ring origin: the gas has been pre-enriched in a galaxy disk and subsequently tidally stripped during a close galaxy-galaxy encounter. 
 
However, opposite to other collisional rings, the Leo ring is not experiencing a vigorous star formation and has an extremely faint optical counterpart. A close analogue, the AGC\,203001 ring \citep{2020MNRAS.492....1B},  has been recently discovered. To better understand the physical conditions driving the nearly quiescent nature of these rings, in this Paper we investigated the young stellar population and the current ability of the Leo ring to form stars. We have four ionized nebulae associated with recent star formation events with far ultraviolet and optical continuum counterparts: two in Clump1, in main body of the ring, and two in Clump2E, a gas droplet in the ring tail towards M\,96. Thanks to HST archival images partially covering Clump1, individual massive stars, or very low mass and compact stellar clusters with a massive outlier, are detected at the center of the nebulae.  

The combination of UV, H$\alpha$ and optical coverage of the star forming regions provides well constrained age and mass estimates for the most massive stars in the ring.  Individual massive stars of O6.5-O7-type might be responsible for the brightest and youngest nebulae in Clump1 and Clump2E. Given the lower gas volume density of the intergalactic ring compared to the ISM in a galaxy, the radius of Str{\"o}mgren spheres expected for the massive stars is indeed compatible with the large size of the \ion{H}{II} regions detected. Massive stars are only a few Myr old, more than 100~Myr younger  than estimated by \citet{2009Natur.457..990T} using only UV and optical continuum. The UV emission, in fact, given its longer decay time and being observed at lower spatial resolution than H$\alpha$, traces star formation for a longer interval of time and over a more extended area, where several sporadic stellar bursts have occurred.  

Star formation in the ring proceeds in high density gas clumps with local burst of 500 -- 1000~$M_\odot$ at intervals of a few Myr forming a rather loose distribution of stars with some massive outliers. The trigger of such bursts can be tidal interactions in the Leo group but also the shock front and feedback from the evolution of  massive stars. In fact, the H$\alpha$ image of Clump2E has revealed a faint H$\alpha$ partial ring hosting a few Myr old \ion{H}{II} region centered around an older, fainter nebula. 

Both Clump1 and Clump2E, as well as other four  \ion{H}{I} clumps which are likely   hosting on-going star formation, follow the Kennicutt-Schmidt relation retrieved for dwarf galaxies and outer disks. For the main body of the ring the global star formation rate  is of order of  $10^{-3}$~$M_\odot$~yr$^{-1}$ or  $10^{-5}$~$M_\odot$~yr$^{-1}$~kpc$^{-1}$ with enhancements close to some \ion{H}{I} peaks and large uncertainties due to stochastic sampling of the IMF.  This  is more than 2 orders of magnitude below that measured in other giant collisional rings such as NGC\,5291 ring \citep{2007A&A...467...93B}.  The extremely long gas depletion time, of order of 100~Gyr, places the ring at the lowest extreme of  the distribution observed in outer disks \citep{2010AJ....140.1194B}.  The metallicity of the Leo ring is only slightly higher than  in outer disks  \citep{2017ASSL..434..145B} but other differences, such as  the greater distance to a star forming disk  and the lack of  gas accretion events (which naturally lead to regions of compressed gas in outer disks),  leave the Leo ring as a unique object to explore. In this environment,  the formation of diffuse metal rich  dwarf galaxies slowly proceeds in the Local Universe. 

Finally we would like to underline the presence of a compact nebula in Clump1 which is a good  planetary nebula candidate. Previous surveys have excluded that planetary brighter  than 27.5~mag in [OIII] exist in the ring. Our candidate planetary nebula  has a brightness close to  this limiting magnitude which gives  $M_{5007}=-2.63^{+0.24}_{-0.20}$~mag,  consistent with the faint optical counterpart which the ring is slowly building up.

\begin{acknowledgements}
Based on observations collected at the European Southern Observatory under ESO program 0104.A-0096(A). EC acknowledges support from PRIN MIUR 2017 -$20173ML3WW_00$ and Mainstream-GasDustpedia. GV acknowledges support from ANID programs FONDECYT Postdoctorado 3200802 and Basal-CATA AFB-170002.  We thank the anonymous referee and L. Magrini for her comments concerning the candidate PNe.
\end{acknowledgements}

%
%

\end{document}